\documentclass[twoside, 10pt, a4paper]{article}

\usepackage[utf8]{inputenc}
\usepackage{amsmath,esint}
\usepackage{graphicx}
\usepackage[colorlinks=true, allcolors=blue]{hyperref}
\usepackage{soul}
\usepackage{subcaption}
\usepackage{amsfonts} 
\usepackage{amsmath,bm}
\usepackage{amssymb}
\usepackage{cancel}
\usepackage{svg}
\usepackage{float}
\usepackage{mathtools}
\usepackage{geometry}
\usepackage{natbib}
\usepackage[affil-it]{authblk}

\newlength{\alphabet}
\title{Spatially localised doubly diffusive convection\\ in an axisymmetric spherical shell}
\author[1]{Paul M. Mannix}
\author[2]{C\'edric Beaume}
\affil[1]{Department of Civil \& Environmental Engineering, Imperial College London, SW7 2AZ, UK}
\affil[2]{School of Mathematics, University of Leeds, LS2 9JT, UK}

\begin{document}
\maketitle
\begin{abstract}
Doubly diffusive convection describes the fluid motion driven by the competition of temperature and salinity gradients diffusing at different rates. While the convective motions driven by these gradients usually occupy the entire domain, parameter regions exist where the convection is spatially localised. Although well-studied in planar geometries, spatially localised doubly diffusive convection has never been investigated in a spherical shell, a geometry of relevance to astrophysics. 
In this paper, numerical simulation is used to compute spatially localised solutions of doubly diffusive convection in an axisymmetric spherical shell. Several families of spatially localised solutions, named using variants of the word convecton, are found and their bifurcation diagram computed. The various convectons are distinguished by their symmetry and by whether they are localised at the poles or at the equator. 
We find that because the convection rolls that develop in the spherical shell are not straight but curve around the inner sphere, their strength varies with latitude, and the system is spatially modulated. As a result spatially periodic states can no longer form and, much like in a planar system with non-standard boundary conditions, localised states are forced to arise via imperfect bifurcations. 
While the direct relevance is to doubly diffusive convection, parallels drawn with the Swift--Hohenberg equation suggest a wide applicability to other pattern forming systems in similar geometries.
\end{abstract}

\section{Introduction}\label{sec:Introduction}

In the oceans, convective transport is driven, in part, by the presence of large-scale gradients of temperature and of salinity. 
As water is heated, its density decreases, resulting in an upward buoyancy force. 
Conversely, when the salt content of a fluid parcel increases, its density increases and yields a downward buoyancy force. 
The fact that these phenomena take place simultaneously and that salt diffuses substantially slower than heat creates a phenomenologically rich physical system referred to as doubly diffusive convection \citep{schmitt94,radko2013double}. 
Due to its importance in sub-planetary scale oceanography, but also as a model problem for pattern formation in fluids, doubly diffusive convection is often studied in a planar fluid layer. 
However, it is also of relevance in a number of astrophysical problems \citep{garaud2018double}.
For example, doubly diffusive instabilities are important flow mechanisms in the Earth's core \citep{trumper2012numerical, monville2019rotating} and doubly diffusive layering has been shown to inhibit vertical transport in the subsurface oceans of Jupiter's and Saturn's icy moons \citep{wong2022layering}. 

\begin{figure}
    \centering
    \includegraphics[width=\textwidth]{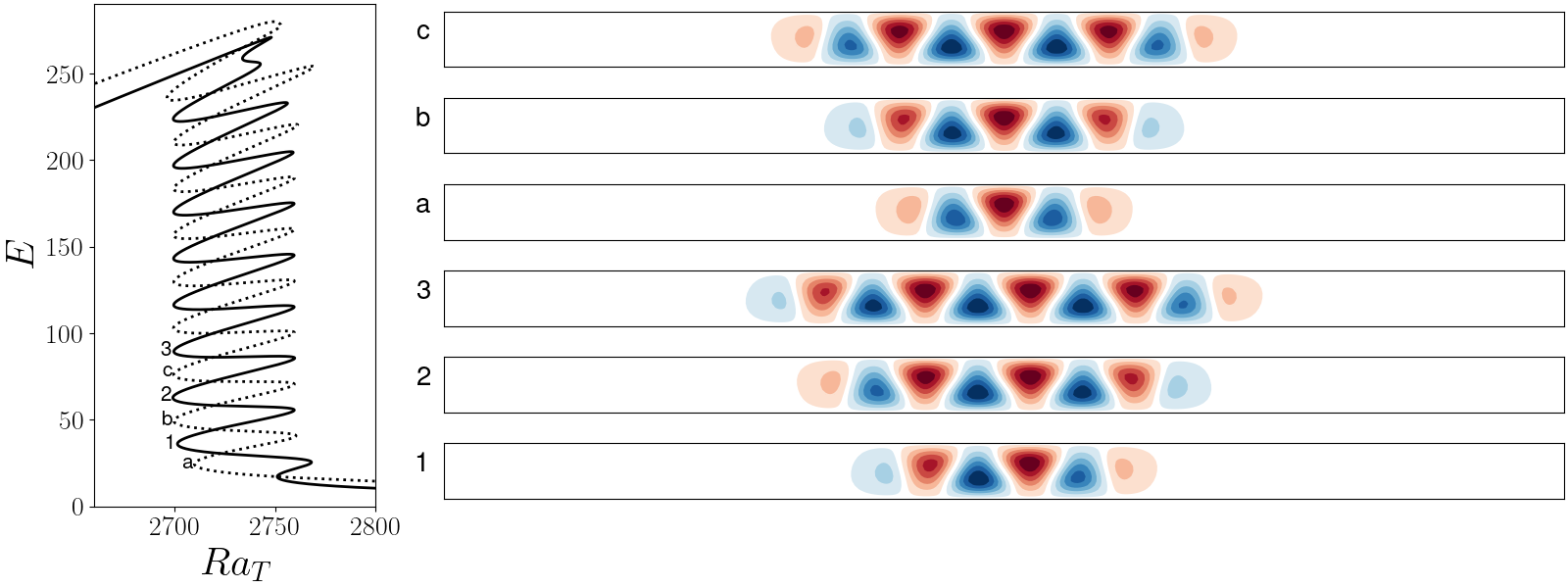}
    \caption{(Left) Bifurcation diagram of the branches of spatially localised convection in planar doubly diffusive convection. The solution branches are represented via their domain-integrated kinetic energy $E$ as a function of the thermal Rayleigh number $Ra_T$. (Right) Representation of the solutions at the saddle-nodes with corresponding labels on the bifurcation diagram. The flow is represented using the temperature deviation/perturbation from the conductive state with red (resp. blue) indicating positive (resp. negative) values. Parameter values are: domain aspect-ratio $20.1643$, solutal Rayleigh number $Ra_S = 500$, Prandtl number $Pr = 1$ and inverse Lewis number $\tau = 1/15$. These results were originally found in \citep{Beaume2011}.}
    \label{fig:planar}
\end{figure}

The initial discovery of steady spatially localised convection consisting of rolls surrounded by large areas of still, conductive fluid \citep{ghorayeb97}, subsequently coined {\it convectons} \citep{blanchflower99}, sparked interest that resulted in the exhaustive characterisation of their bifurcation diagram in natural doubly diffusive convection, i.e., in vertically extended domains where convection is driven by horizontal gradients of temperature and salinity \citep{bergeon2008periodic,beaume2013convectons}.
Convectons were also studied in binary fluid convection (the doubly diffusive convection that takes place in a horizontal fluid layer driven by vertical gradients) in the presence \citep{mercader2009localized,mercader2011convectons} and in the absence \citep{Beaume2011} of Soret effect.
The bifurcation diagram in the latter case shows two branches of convectons which we recomputed in figure \ref{fig:planar}. 
The solutions along these branches posses different parity, with either an odd or even number of rolls, respectively corresponding to an even or an odd number of local extrema of the temperature perturbation profile.
The solution branches neatly organise themselves to undergo a series of saddle-node bifurcations between two well-defined values of the thermal Rayleigh number. 
This behaviour is called {\it snaking} and is accompanied with a spatial growth mechanism, whereby a convecton nucleates one roll on either of its sides every time the branch is continued upward past two saddle-nodes.
This can be observed by comparing successive left saddle-node states in figure \ref{fig:planar}. 
An important consequence of the snaking structure is that, for a given range of parameter values, a large number of spatially localised states of different sizes coexist. 
For an exhaustive description of the formation of spatially localised structures, we refer the reader to the recent reviews by \citep{knobloch2015spatial} and by \citep{bramburger25arxiv} and to early work on homoclinic snaking in the Swift--Hohenberg equation \citep{burke06}.

Following \citep{mercader2011convectons} and \citep{Beaume2011,beaume2013convectons,beaume2013nonsnaking}, a good understanding of the formation of spatially localised structures in doubly diffusive convection, for simple planar geometries, is now available. 
For this reason, recent effort has veered away from the idealised setups initially used. 
The effect of spatial symmetry breaking by a change in the solutal boundary condition in natural doubly diffusive convection revealed that convectons became asymmetric and started travelling \citep{lojacono17}. 
Spatial symmetry breaking was also explored in binary fluid convection through the inclination of the horizontal fluid layer which resulted in dramatic changes to the solution space, including the discovery of a new type of spatially localised solution \citep{mercader19,alonso2022stationary}.
Other work investing the role of temporal symmetry has focused on travelling localised states and their interaction \citep{watanabe2012,watanabe2016}. 
Beyond understanding the role of symmetries, recent investigations into natural doubly diffusive convection have also sought to understand the importance of subcriticality as well as that of the balance between thermal and solutal effects. 
In particular, \citep{tumelty23} revealed the persistence of spatially localised states into the supercritical regime, albeit with an altered bifurcation scenario, while \citep{tumelty2025} explored 
the formation of spatially localised states away from the balanced case where thermal and solutal effects are equally important, 
identifying the emergence of convectons via cusps in the thermally dominated regime and a smooth connection to domain-filling convection in the solutally dominated one.
Understanding how robust convectons are to such departures from the idealised setups in which they were initially investigated is fundamental to transfer knowledge to less ideal systems.

It is now natural to turn to other geometries in order to understand spatial localisation more generally.
Although the formation of spatially localised states in cylindrical geometries has been investigated for the Swift--Hohenberg equation \citep{mccalla_snaking_2010, mccalla_spots_2013, verschueren_localized_2021}, to the authors' knowledge, it has yet to be studied in a spherical geometry.
The absence of such studies is however not withstanding the fact that examples of spherical physical systems susceptible to harbouring such states do exist. 
For example, the wrinkling of elastic rings has recently been investigated to show the emergence of spatially localised folds \citep{foster22} and fingering patterns \citep{foster23}, and is therefore suggestive of the existence of a spatial localisation mechanism in the spherical extension of the problem. 
Similarly, the wrinkling of spherical membranes indeed shows a hysteretic transition between the unpatterned surface and hexagons \citep{stoop2015curvature}, much like the scenario observed in the Rosensweig instability that led to the discovery of snaking in magnetically forced ferrofluids \citep{Lloyd15}. 
In an astrophysical context, the geodynamo is believed to originate via a subcritical bifurcation \citep{rincon_dynamo_2019} and, at its laminar to turbulent boundary in phase space, also admits unstable travelling wave solutions \citep{skene2024nonlinear}. 
Given the role played by localised flow states in transition to turbulence in shear flows \citep{duguet09, schneider2010snakes, Gibson2016, pershin19}, it is possible that similar states could help to understand the transition to magnetohydrodynamic (MHD) turbulence in spherical geometries.

To understand the influence of the geometry on spatial localisation in fluids, this paper asks: Do spatially localized solutions of the Navier--Stokes equations exist for convection in a spherical shell?
And, if so, what do these solutions look like and how do they form? 
In order to enable useful links with established literature, we consider doubly diffusive convection and use the same set up as in \citep{Beaume2011} with the only change being the geometry: in previous work, a two-dimensional planar fluid layer with spatially periodic boundary conditions was used, while we are posing the equations and boundary conditions on an axisymmetric spherical shell.

In section \ref{sec:Formulation} the problem formulation is presented. 
We then introduce, in section \ref{sec:Solutions}, the types of solution we found in order to provide their spherical representation and establish terminology. 
We go back to the basics in section \ref{sec:LinearProb}, where we solve the linear stability problem for the conduction state, followed, in section \ref{sec:Results}, by the presentation of our main results. Finally, this paper terminates in section \ref{sec:Discussion} by a discussion.

\section{Formulation}\label{sec:Formulation}

We consider convection driven by competing buoyancy forces generated by temperature and solute concentration gradients between spheres of radii $R_1$ and $R_2>R_1$. 
The forcing gradients are imposed via a constant temperature and solute concentration at the sphere walls.
The inner sphere is set to be hotter and richer in solute such that the temperature and solute concentration differences are $\Delta T = T_1 - T_2 > 0$ and $\Delta S = S_1 - S_2 > 0$, where $T_1$ and $S_1$ (resp. $T_2$ and $S_2$) are taken at the wall of the sphere of radius $R_1$ (resp. $R_2$).
We assume that the fluid contained within the resulting spherical shell is well-modelled under the Boussinesq approximation: its density is taken to be $\rho_f$ everywhere except in the gravitational force term, where it varies linearly with the fluid temperature and solute concentration:
\begin{equation}
\rho(T,S) = \rho_f [1 - \beta_T(T - T_2) + \beta_S (S - S_2) ].
\end{equation} 
Here, the expansion coefficients $\beta_T$ and $\beta_S$ are 
positive constants and $\rho_f$ is the density at the reference temperature $T_2$ and solute concentration $S_2$.
The gravity field is spherically symmetric and follows from the assumption that the inner core ($r<R_1$) density is much greater than that of the spherical shell fluid:
\begin{equation}
    \boldsymbol{g} = - g_0 \frac{R_1^2}{r^2} \boldsymbol{\hat{r}},
\end{equation}
where $g_0$ is the acceleration due to gravity on the inner sphere surface. 
Choosing the length scale $d = R_2 - R_1$, the thermal diffusion time scale $d^2/\kappa$ and the temperature and solutal scales $\Delta T$ and $\Delta S$, the governing equations may be written in the following nondimensional form:
\begin{eqnarray}
\frac{1}{Pr} \left( \frac{D \boldsymbol{u}}{D t} + \nabla P \right) &=& g(r) \hat{\boldsymbol{r}} (Ra_T T  -  Ra_S \, S)  + \nabla^2 \boldsymbol{u},
\label{eq:momentum_3D}\\
\frac{\partial T}{\partial t} + \boldsymbol{u} \cdot \nabla T  &=& \nabla^2 T,
\label{eq:Temp_3D}\\
\frac{\partial S}{\partial t} + {\bf u} \cdot \nabla S &=&  \tau \nabla^2 S,
\label{eq:solutal_3D}\\
\boldsymbol{\nabla} \cdot \boldsymbol{u} &=& 0,
\label{eq:Cont_3D}
\label{eq:3D_EQNS}
\end{eqnarray}
where $r$ is now nondimensional, ${\boldsymbol u}$ is the velocity field, $P$ is the pressure, $T$ is the temperature, $S$ is the solute concentration and $t$ is the time.
In writing these equations, we have introduced a number of parameters:
\begin{eqnarray}
    \text{Prandtl number:~} &Pr =& \displaystyle\frac{\nu}{\kappa},\\
    \text{Lewis number:~} &\tau =& \displaystyle\frac{\kappa_S}{\kappa},\\ 
    \text{(Thermal) Rayleigh number:~} &Ra_T =& \displaystyle\frac{g_0 \beta_T \Delta T d^3}{\nu \kappa} > 0,\\
    \text{Solutal Rayleigh number:~} &Ra_S =& \displaystyle\frac{g_0 \beta_S \Delta S d^3}{\nu \kappa} > 0,\\
    \text{Aspect ratio:~} &\Gamma =& \displaystyle\frac{R_1 \pi}{d},\\
    \text{Gravitational factor:~} &g(r) =& \displaystyle\frac{R_1^2}{r^2 d^2},
\end{eqnarray}
where $\nu$ is the kinematic viscosity, $\kappa$ the thermal diffusivity and $\kappa_S$ the solutal diffusivity.
The thermal and solutal Rayleigh numbers measure the strength of the buoyancy force resulting from the imposed temperature and solutal gradients at the sphere boundaries, the Prandtl and inverse Lewis numbers determine the dominant diffusive processes while $\Gamma$ and $g(r)$ are geometric parameters.
We specify no-slip velocity boundary conditions for $\boldsymbol{u}$ and assume perfectly conducting spherical shell walls
\begin{align}
\boldsymbol{u}(r=r_1,\theta,\varphi) &= 0, \; \boldsymbol{u}(r=r_2,\theta,\varphi) = 0, \\
T(r=r_1,\theta,\varphi) &= 1, \; T(r=r_2,\theta,\varphi) = 0, \\
S(r=r_1,\theta,\varphi) &= 1, \; S(r=r_2,\theta,\varphi) = 0,
\label{eq:bcs_uTS}
\end{align}
where $r_1 = R_1/(R_2 - R_1)$ (resp. $r_2 = R_2/(R_2 - R_1)$) is the location of the inner (resp. outer) sphere wall, $\theta$ is the polar coordinate and $\varphi$ is the azimuthal coordinate.
We look for longitudinally invariant solutions that are devoid of zonal flow and therefore impose $\partial/\partial \varphi = 0$ and $u_\varphi = 0$.
This restriction allows the use of the following pole boundary conditions: 
\begin{equation} 
\frac{\partial S}{\partial \theta}  = \frac{\partial T}{\partial \theta} = \frac{\partial u_r}{\partial \theta} = u_{\theta} = 0, \quad \text{for} \quad \theta = 0,\pi,
\label{eq:SYMM_CONDS}
\end{equation}
where $\theta = 0$ (resp. $\pi$) can be thought of as the North (resp. South) pole of our spherical geometry. 

The problem solved by \citep{Beaume2011} in a planar geometry, and whose results were reproduced in figure \ref{fig:planar}, is different from the one we pose on the spherical shell in three important ways: (i) the system no longer has the top-down reflection symmetry of the plane layer, (ii) the geometry is curved as opposed to flat and (iii) the $\theta$ boundary conditions are of a different nature than the periodic boundary conditions used in the planar problem.

System \eqref{eq:3D_EQNS}--\eqref{eq:SYMM_CONDS} admits the base state solution
\begin{equation} 
\boldsymbol{u} = 0, \quad T_0 = S_0 = - A_T/r + B_T,
\label{eq:conductive_state}
\end{equation}
where $A_T = \frac{r_1 r_2}{r_1 - r_2}$ and $B_T = \frac{r_1}{r_1 - r_2}$. To understand the symmetries of the system, we introduce the convective variables $\Theta$ and $\Sigma$ such that $T = T_0 + \Theta$ and $S = S_0 + \Sigma$. 
As discussed by \cite{beltrame2015onset}, the full three-dimensional system is equivariant under the antipodal symmetry $\boldsymbol{r} \to -\boldsymbol{r}$.
Using longitudinal invariance, this symmetry reads:
\begin{equation}
    R_{AP}: (r,\theta) \to (r, \pi - \theta), \quad (u_r, u_{\theta}, \Theta, \Sigma) \to (u_r, -u_{\theta}, \Theta, \Sigma)
    \label{eq:symmetry}
\end{equation}
and can be understood as a simple reflection about the equator. 
Spatial reversibility, here associated with $R_{AP}$, is an important feature of systems that admit stationary spatially localised solutions \citep{Burke2009}.
For example, if a solution is conductive at the North pole and displays convection rolls at the equator, then the reverse connection can be found from the equator to the South pole, yielding a spatially localised convection structure at the equator. 

In practice, the fact the flow is two-dimensional and incompressible allows the introduction of a streamfunction $\psi$: 
\begin{equation}
\boldsymbol{u} = \nabla \times \left(\frac{\psi}{r} \boldsymbol{\hat{\varphi}}\right) = \bigg( \frac{1}{r^2 \sin \theta} \frac{\partial (\psi \sin \theta)}{\partial \theta}, - \frac{1}{r}\frac{\partial \psi}{\partial r}, 0 \bigg),
\end{equation}
which is chosen following \cite{Marcus1987} so that the resulting primitive variable system projects onto a sine/cosine basis. Using this definition of the streamfunction, the primitive variable system \eqref{eq:3D_EQNS} can then be written as
\begin{subequations}
\begin{equation}
\frac{\partial D^2 \psi }{\partial t}  + \mathcal{J}( \psi, \psi) = Pr g(r) \bigg( Ra_T \frac{\partial \Theta}{\partial \theta} -  Ra_S \frac{\partial \Sigma}{\partial \theta} \bigg) + Pr D^2 D^2 \psi,
\label{eq:PDE_Vorticity}
\end{equation}
\begin{equation}
\frac{\partial \Theta}{\partial t} + J( \psi, T_0 + \Theta ) = \nabla^2 \Theta,
\label{eq:PDE_Thermal}
\end{equation}
\begin{equation}
\frac{\partial \Sigma}{\partial t} +J( \psi, S_0 + \Sigma) =  \tau \nabla^2 \Sigma,
\label{eq:PDE_Solute}
\end{equation}
\label{eq:Full_Eqs_App}
\end{subequations}
where
\begin{equation}
J(\psi,f) = \frac{1}{r^2 \sin \theta} \frac{\partial (\psi \sin \theta) }{\partial  \theta} \frac{\partial f }{\partial  r} - \frac{1}{r^2} \frac{\partial \psi}{\partial r} \frac{\partial f }{\partial  \theta},
\end{equation}
\begin{equation}
\mathcal{J}( \psi, \tilde{\psi}) = \frac{\partial }{\partial r} \bigg( ( r^2 u_r) \frac{D^2 \tilde{\psi}}{r^2} \bigg) + \frac{ \partial }{\partial \theta} \bigg( (r u_{\theta}) \frac{D^2 \tilde{\psi}}{r^2} \bigg)
\end{equation}
and
\begin{equation}
\nabla^2 = \frac{1}{r^2} \frac{\partial }{\partial r} \bigg( r^2 \frac{\partial }{\partial r} \bigg) + \frac{1}{r^2 \sin \theta} \frac{\partial}{\partial \theta} \bigg(\sin \theta \frac{\partial }{\partial \theta} \bigg), 
\end{equation}
\begin{equation}
D^2 = \frac{\partial^2 }{\partial r^2} + \frac{1}{r^2 \sin \theta} \frac{\partial}{\partial \theta}\bigg(\sin \theta \frac{\partial }{\partial \theta} \bigg) - \frac{1}{r^2 \sin^2 \theta}.
\end{equation}
In this formulation, the spherical shell boundary conditions become
\begin{equation}
 \psi = \frac{\partial \psi}{\partial r} = \Theta= \Sigma = 0, \quad \hbox{for} \; \; r = r_1,r_2,
\label{eq:bcs_App}
\end{equation}
while the pole conditions are
\begin{equation} 
\psi = \frac{\partial \Theta}{\partial \theta} = \frac{\partial \Sigma}{\partial \theta}=0, \quad \text{for} \quad \theta = 0,\pi.
\end{equation}

Following \citep{Beaume2011}, which studied the formation of spatially localized convection states in doubly diffusive convection over a periodic planar geometry, we set $\tau = 1/15$ and $Pr = 1$. 
We use the thermal Rayleigh number as the control parameter against which the bifurcation diagrams of steady states will be represented.
However, to enable us to fully explain the localized pattern formation at play in our system, we have had to vary also the solutal Rayleigh number.

\section{Solutions}\label{sec:Solutions}

We now outline the main types of localised solutions discussed in this paper.
We concentrate exclusively on the simplest types of localised solutions constituting straightforward homoclinic or heteroclinic connections in space. 
More complex connections exist but are beyond the scope of our work.

\begin{figure}
    \centering
    \includegraphics[width=0.95\textwidth]{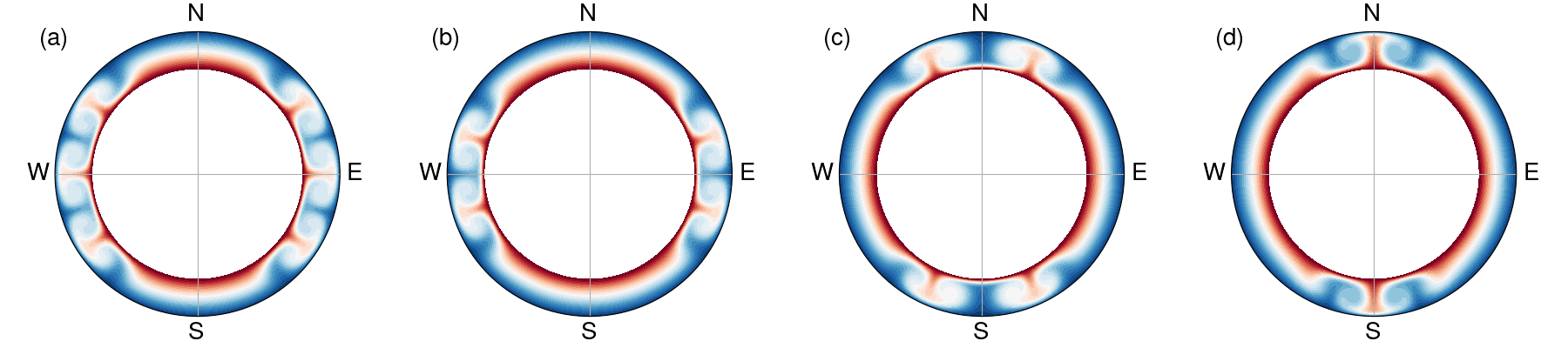}
    \caption{Equatorially-symmetric spatially localised states of convection represented by the salinity (red for saltier) in a section of the spherical shell passing by the poles. Four different types of solutions are shown: (a) $C+$ equatorial-convectons, (b) $C-$ equatorial-convectons, (c) $A+$ anticonvectons and (d) $A-$ anticonvectons. 
    These states are computed at $Ra_S = 400, \Gamma = 8.9224$ for different $Ra_T$ and are discussed further in \autoref{sec:even_results}.
    } 
    \label{fig:Intro_Even_Parity}
\end{figure}

The first family of spatially localised solutions consists of solutions that respect the symmetry $R_{AP}$, i.e., their reflection about the equator returns the same solution.
Examples of such solutions are shown in terms of the solutal field $S$ in figure \ref{fig:Intro_Even_Parity}.
Since they display identical variations in salinity (and in temperature) on either side of the equator, they can be understood as even parity states with respect to the equator.
Consequently, no convection roll can straddle the equator and, instead, a pair of counter-rotating rolls is present and separated by the equatorial plane on which a purely vertical (i.e., aligned with gravity) flow is observed.
We found two types of $R_{AP}$-symmetric localised states.
Solutions represented in panels (a) and (b) of figure \ref{fig:Intro_Even_Parity} are characterised by the presence of convection around the equator while the fluid near the poles is mostly still.
We will refer to these states as {\it equatorial-convectons}.
The convecton of panel (a) displays an upward flow at the equator and will be referred to as $C+$.
Conversely, the convecton of panel (b) will be referred to as $C-$ due to the downward flow it produces at the equator.
We have also identified localised solutions where the convection is centred around both poles and the equatorial region displays mostly quiescent fluid.
Examples of such states are shown in panels (c) and (d).
Owing to the imposed longitudinal invariance, the poles create a constraint on the flow and they bear similarity with solutions found in planar geometry with end-walls, where they are known as {\it anticonvectons} \citep{mercader2011convectons, tumelty2025}.
We will adopt this terminology here.
The anticonvectons in panel (c) connect with the $C+$ convectons (panel(a)), as shown in section \ref{sec:even_results}, so we will refer to them as $A+$ anticonvectons. 
A similar connection is found between $C-$ convectons (panel(b)) and the anticonvectons of panel (d), which will be referred to as $A-$ anticonvectons.

\begin{figure}
    \centering
    \includegraphics[width=0.5\textwidth]{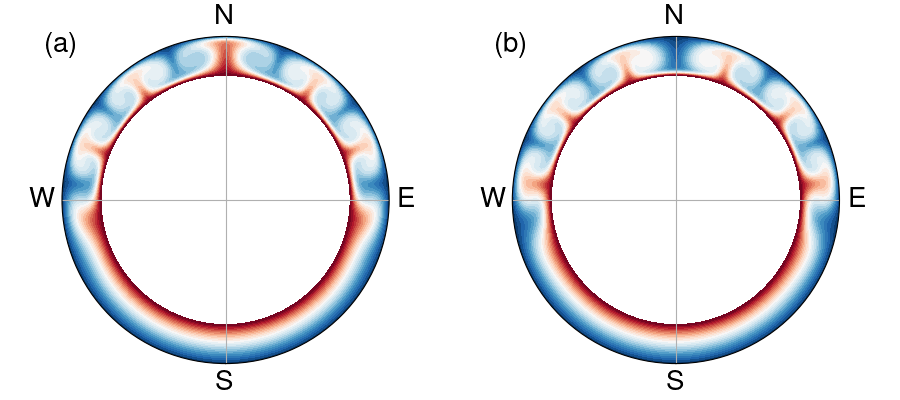}
    \caption{Symmetry-breaking spatially localised states of convection represented by the salinity (red for saltier) in a section of the spherical shell passing by the poles. Two different types of solutions are shown: (a) $+$ pole-convecton and (b) $-$ pole-convecton. 
    These states are computed at $Ra_S = 150, \Gamma = 10.029$ for different values of $Ra_T$ and are discussed further in \autoref{sec:odd_results}.} 
    \label{fig:Intro_Odd_Parity}
\end{figure}

In addition to symmetric convectons, we are also interested in spatially localised solutions that break $R_{AP}$.
The simplest such solutions are shown in figure \ref{fig:Intro_Odd_Parity} and consist of states that display convection near one pole only.
We call these {\it pole-convectons}.
We found two families of such states, which we differentiate by the direction of the flow at the pole.
We will refer to those with an upward flow at the pole as $+$ pole-convectons (figure~\ref{fig:Intro_Odd_Parity}(a)), while those with a downward flow at the pole will be referred to as $-$ pole-convectons (figure~\ref{fig:Intro_Odd_Parity}(b)).
Similar states consisting of convection near the South pole and conduction near the North pole also exist but will not be reported here: they can be obtained by applying the broken symmetry, $R_{AP}$, onto the solutions presented and are, consequently, dynamically equivalent.
Pole-convectons share a similar spatial structure with localised wall states found in planar binary fluid convection \citep{mercader2011convectons}.

Before discussing how these six types of solutions arise, we briefly comment on the consequences of imposing longitudinal-invariance. 
As made apparent by figures \ref{fig:Intro_Even_Parity} and \ref{fig:Intro_Odd_Parity}, an axis of symmetry passing through the sphere's North and South poles now exists.
The existence of this axisymmetry, in addition to making computations easier and simplifying phase space, implies that convection rolls near the poles take a doughnut- or torus-like form and convect fluid away from the spherical shell boundaries over a smaller surface area than a convection roll of a similar cross section located near the equator. 
To simplify visualisation, we will present our solutions on a plane where the South pole can be thought to be at the left end and the North pole at the right end. 
While this geometrical feature is easy to visualise in Figures \ref{fig:Intro_Even_Parity} and \ref{fig:Intro_Odd_Parity}, it is not that clear on the flattened representations that follow. 
The reader should therefore keep in mind the physical differences between pole (outer regions in the flattened figures) and equatorial rolls.

\section{Linear Theory}\label{sec:LinearProb}

To solve the linear problem, we chose a different spherical streamfunction formulation:
\begin{equation}
\boldsymbol{u} = \nabla \times \left(0,0,\frac{\Psi(r,\theta,t)}{r \sin \theta}\right) = \bigg( \frac{1}{r^2 \sin \theta}\frac{\partial \Psi}{\partial \theta}, \frac{-1}{r \sin \theta}\frac{\partial \Psi}{\partial r}, 0 \bigg),
\label{eq:Stream_Function_analytical}
\end{equation}
such that the linear system admits separable solutions in $r$ and $\theta$ \citep{beltrame2015onset}. Upon using this streamfunction in system \eqref{eq:3D_EQNS} and linearising the resulting equations about the conduction state \eqref{eq:conductive_state}, one obtains the perturbation system:
\begin{equation}
\bigg( \frac{\partial }{\partial t} \mathcal{M} - \mathcal{L} \bigg) \boldsymbol{\tilde{X}} = 0, \quad \boldsymbol{\tilde{X}} = \begin{pmatrix} \tilde{\Psi} & \tilde{\Theta} &  \tilde{\Sigma} \end{pmatrix}^T,
\label{eq:Mat_LIN}
\end{equation}
where
\begin{equation}
\mathcal{M} =  \begin{pmatrix} \displaystyle\frac{1}{Pr} \, \hat{D}^2 & 0 & 0\\ 0 &  \mathbb{I} & 0 \\ 0 & 0 &  \mathbb{I} \end{pmatrix}, \quad
\mathcal{L} =   \begin{pmatrix} \hat{D}^2 \hat{D}^2 & Ra_T \, g(r) \sin \theta \displaystyle\frac{\partial }{\partial \theta} & - Ra_S \, g(r) \sin \theta \displaystyle\frac{\partial }{\partial \theta},\\ -\displaystyle\frac{T'_0}{r^2 \sin \theta} \frac{\partial}{\partial \theta} &  \nabla^2 & 0 \\ -\displaystyle\frac{S'_0}{r^2 \sin \theta} \frac{\partial}{\partial \theta} & 0 & \tau \nabla^2 \end{pmatrix},
\label{eq:linearoperator}
\end{equation}
and $\hat{D}^2 = D^2 + \frac{1}{r^2 \sin^2 \theta}$. 
In writing \eqref{eq:linearoperator}, we have introduced the radial derivatives of the base state: $T_0' = \partial T_0 / \partial r$ and $S_0' = \partial S_0 / \partial r$. 
Owing to the choice of streamfunction \eqref{eq:Stream_Function_analytical}, system \eqref{eq:Mat_LIN} admits separable eigenfunctions of the form
\begin{eqnarray}
\tilde{\Psi} &=& e^{\lambda t} \, \Psi_{\ell}(r) \, \sin \theta \frac{\partial P_{\ell}(\cos \theta)}{\partial \theta},\\
\tilde{\Theta} &=& e^{\lambda t} \, \Theta_{\ell}(r) \, P_{\ell}(\cos \theta),\\
\tilde{\Sigma} &=& e^{\lambda t} \, \Sigma_{\ell}(r) \, P_{\ell}(\cos \theta).
\label{eq:lin_sols}
\end{eqnarray}
In writing these eigenfunction, we have introduced the temporal growth rate $\lambda \in \mathbb{C}$ as well as $P_{\ell}$, the Legendre polynomial of degree $\ell$. 
Substituting for $\Psi_{\ell}(r), \Theta_{\ell}(r), \Sigma_{\ell}(r)$ and using the fact that $T'_0(r) = S'_0(r)$ allows System \eqref{eq:Mat_LIN} to be reduced to an $8^{th}$ order ODE:
\begin{equation}
( \lambda  - \tau \nabla^2_{\ell}) (\lambda  - \nabla^2_{\ell}) g(r)^{-1}  \left(  \frac{\lambda}{Pr} D_{\ell}^2 - D_{\ell}^2 D_{\ell}^2 \right) \Psi_{\ell} = \left( \lambda - \nabla^2_{\ell} \right) \tilde{Ra} \frac{T'_0}{r^2} \ell(\ell+1) \Psi_{\ell},
\label{eq:ode_for_l_general}
\end{equation}
where $\tilde{Ra} = Ra_T - Ra_S$ and
\begin{equation}
D^2_{\ell} = \frac{\partial^2 }{\partial r^2} - \frac{\ell(\ell + 1) }{r^2}, \quad
\nabla^2_{\ell} = \frac{1}{r^2} \frac{\partial }{\partial r} \bigg( r^2 \frac{\partial }{\partial r} \bigg) -  \frac{\ell(\ell +1)}{r^2}.
\end{equation}

\begin{figure}
    \centering
    \includegraphics[width=0.95\textwidth]{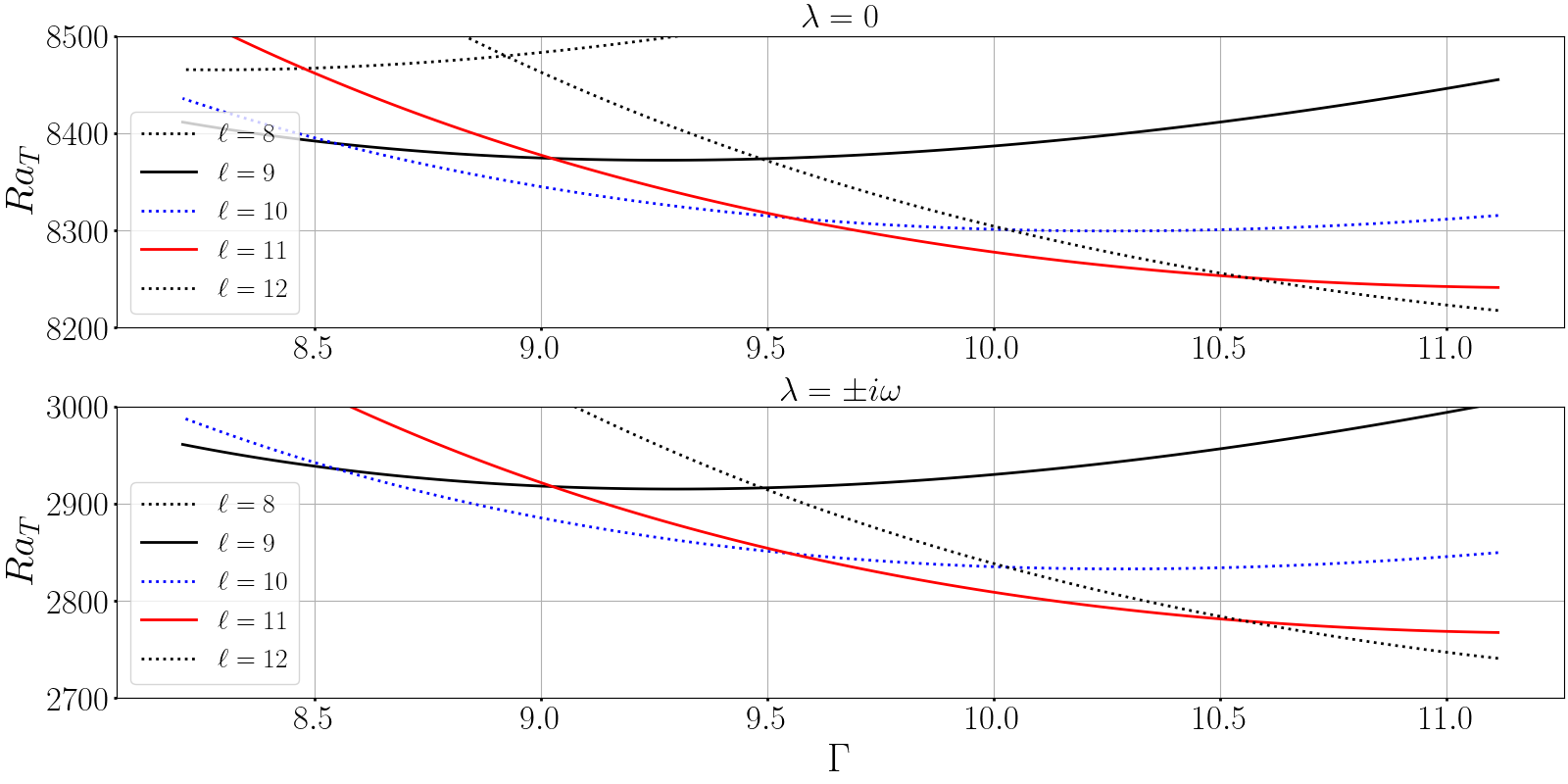}
    \caption{Neutral curves demarcating the onset of convection via steady state ($\lambda = 0)$ and oscillatory state ($\lambda = \pm i \omega$) bifurcations, where $\ell$ denotes the Legendre polynomial associated with the latitudinal mode. These curves were computed for $Ra_S=400$. According to equation \eqref{eq:ode_for_l_general}, the critical value of $Ra_T$ varies trivially with $Ra_S$ as the critical value of $Ra_T - Ra_S$ remains constant. }
    \label{fig:Neutral_Curves}
\end{figure}

Letting $\lambda = s \pm i \omega$ and setting $s=0$, one can solve for the critical $\tilde{Ra}$ for a given aspect ratio $\Gamma$ and Legendre polynomial degree $\ell$. 
Figure \ref{fig:Neutral_Curves} shows the resulting neutral stability curves as the aspect ratio $\Gamma$ varies for fixed $Ra_S$.  
The advantage of the above method is that the Legendre polynomials in $\theta$ are decoupled, allowing to track straightforwardly the neutral stability curve of the various latitudinal modes. 
As $Ra_T$ increases from zero, the conduction state first loses stability to time-dependent convection via a Hopf bifurcation, indicated by the neutral stability curves for which $\lambda = \pm i \omega$. 
Bifurcations to steady states only come in at larger values of the thermal Rayleigh number.

Although we present results between these two types of bifurcations separately, the dynamics they generate may impact one another. 
When posed on a flat plane, two-dimensional doubly diffusive convection possesses the symmetries of the group $O(2) \times Z_2$ which leads to simultaneous emergence of standing and travelling waves \citep{Crawford91}. 
For the values of the parameters we chose, \citep{Beaume2011} found that the travelling waves bifurcate subcritically but soon turn around a saddle-node and extend to larger values of the thermal Rayleigh number until they connect with a branch of spatially periodic steady convection, to which they transfer their stability. 
On the one hand, the equations posed in a spherical shell are not symmetric with respect to horizontal translations or vertical reflections. 
They simply possess the symmetry $Z_2 = \{ I, R_{AP}\}$ and Hopf bifurcations will only create standing waves. 
On the other hand, the larger $\Gamma$, the more the present spherical system approaches the planar system of \citep{Beaume2011}. 
In the planar problem, steady state bifurcations are subcritical and give rise to convectons, which is the observation that forms the basis of our investigation here.

\begin{figure}
\centering
\includegraphics[width=\textwidth]{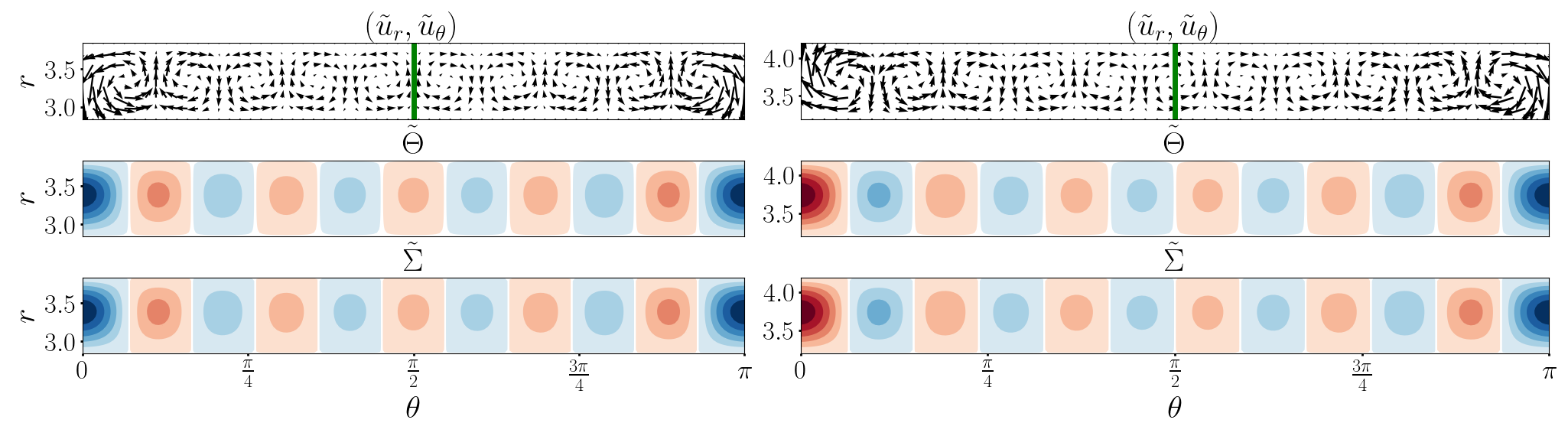}
\caption{(Left) Even parity mode $\ell = 10, \Gamma = 8.9224$ and (Right) odd parity mode $\ell = 11, \Gamma = 10.029$ at onset represented by the perturbations (top panels) in the in-plane velocity, (middle panels) in the temperature and (bottom panels) in the salinity fields. The velocity is shown by arrows indicating the magnitude and direction of the flow while positive (resp. negative) values of the temperature and salinity perturbations shown in red (resp. blue). The even parity mode respects the $R_{AP}$ symmetry while the odd parity one breaks it. Both modes are spatially modulated such that their maximum occurs at the poles and minimum at the equator.}
\label{fig:EigenVectors}
\end{figure}

We decided on the value of the aspect ratio $\Gamma$ in the following way. 
First of all, we note that equatorial-convectons and anticonvectons respect $R_{AP}$ and are generated by even Legendre modes. 
We will be studying them in the subspace generated by the even latitudinal modes. 
We set $\ell = 10$ to enable a sufficiently large number of convection rolls to develop and fixed $\Gamma = 8.9224$. 
This value was chosen so that the next even-parity steady state bifurcations are the farthest away: for $Ra_S = 400$, the $\ell = 10$ bifurcation is located at $Ra_{T,10} \approx 8351.53$ while the bifurcations to $\ell = 8$ and $\ell = 12$ modes only occur at $Ra_{T,8} \approx Ra_{T,12} \approx 8479.92$. 
Pole-convectons break the symmetry and their search will require the full solution space. 
We set $\ell = 11$ in order to enable the formation of structures that break $R_{AP}$. 
We chose $\Gamma = 10.029$, the value of the aspect ratio for which the next steady state bifurcations (toward $\ell = 10$ and $\ell = 12$) are the furthest away. 
The steady state bifurcation to $\ell = 11$ modes is found at $Ra_{T,11} \approx 8275.70$ for $Ra_S = 400$. The $\ell = 10$ and $\ell = 11$ eigenvectors that are responsible for these steady state bifurcations are shown in figure \ref{fig:EigenVectors}.
Due to its parity, the $\ell = 10$ mode generates either an upward or a downward flow at both poles as well as a flow in the opposite direction at the equator. 
Conversely, the $\ell = 4n$ modes, where $n = 1, 2, \dots$, trigger convective structures that display flows in the same direction at both poles and over the equator. 
The even modes preserve $R_{AP}$ and, thus, generate what may be thought of as transcritical bifurcations. 
Like all odd modes, the $\ell = 11$ mode is characterized by an equator-straddling roll and by opposite vertical flows at the poles. 
They break the $R_{AP}$ symmetry and are thus responsible for the occurrence of pitchfork bifurcations of the conduction state.

In addition to the symmetry group $Z_2$, System \eqref{eq:3D_EQNS} also possesses a hidden symmetry due to the operator in equation \eqref{eq:Mat_LIN} being self-adjoint. 
As discussed by \cite{beltrame2015onset}, this hidden symmetry, here due to $g(r) \sim r^{-2}, T'_0(r) \sim r^{-2}, S'_0(r) \sim r^{-2}$, implies that the bifurcation to even parity solutions is a codimension-2 bifurcation corresponding to the collision between the transcritical bifurcation from the conduction state and the saddle-node bifurcation of the supercritically emerging branch. 
We will refer to this transcritical-saddle-node bifurcation as a TSN bifurcation. 
The unfolding of this bifurcation can be carried out by varying the functional form of $g(r)$ relative to $T'_0(r)$ and $S'_0(r)$ and is out of the scope of this paper.

Due to the spherical geometry, the eigenfunctions in the linear stability analysis consist of a spatially modulated array of convection rolls, where the polar rolls are the strongest (see figure \ref{fig:EigenVectors}). 
This spatial modulation is not a consequence of the pole boundary conditions \eqref{eq:SYMM_CONDS}, which can be thought of as Neumann boundary conditions, but rather because of the presence of sinusoidal terms in the governing equations. 
Physically, this modulation occurs because the contact plane between rolls near the poles has a smaller surface area than near the equator and, thus, rolls have to be stronger to transfer the same amount of heat or solute.
This is not a consequence of axisymmetry but rather of the fact that convection rolls on a sphere are necessarily curved.

Further insight can be obtained by drawing analogies between the work of \cite{Lloyd09}, who studied the Swift--Hohenberg equation on an axisymmetric disc, and the spherical shell system studied here in the limit of large $\Gamma$. 
The similarities between these systems arise from the fact that we look for solutions that are axisymmetric and, thus, that the spatial dynamics near the pole resemble that of the axisymmetric disc near its origin.
The eigenfunctions of the linear stability analysis around the trivial state of the axisymmetric disc problem are Bessel functions which decay radially while oscillating.
In the present spherical shell problem, the counterpart eigenfunctions are Legendre polynomials whose modulus is maximal at the poles.
In the limit of large $\Gamma$, these Legendre polynomials are closely related to Bessel functions, further emphasizing the connection between the spatial modulation observed here and that studied in the axisymmetric disc.
We therefore anticipate that, in the limit of large $\Gamma$, anticonvectons (see figure \ref{fig:Intro_Even_Parity}) and pole-convectons (see figure \ref{fig:Intro_Odd_Parity}) will have a similar bifurcation scenario to that of the spot-like localised states bifurcating from the zero state 
in the Swift--Hohenberg equation \citep{Lloyd09}.

\section{Results}\label{sec:Results}

Following \cite{Marcus1987}, we solve system \eqref{eq:Full_Eqs_App} using a pseudo-spectral method that employs either a sine basis or cosine basis in $\theta$ together with a Chebyshev basis in $r$. 
We perform differentiation in collocation space for the radial direction and dealiase nonlinear terms using the 3/2-rule for latitudinal modes such that the largest resolved wavenumber is $N_{\theta}$. 
Our numerical scheme relies on the use of transforms to send physical fields into spectral space. 
These transforms have $O(N_r^2 N_{\theta}\log(N_{\theta}))$ numerical complexity, where $N_{\theta}$ (resp. $N_r$) refers to the number of sine/cosine modes (resp. Chebyshev collocation points) used. We perform time-stepping, steady-state computation and numerical continuation using matrix free methods following \citep{Tuckerman1989,Edwards1994,beaume2017adaptive,Tuckerman2019}.
Our code is validated against the Dedalus code \citep{burns2020dedalus} and is available on \href{https://github.com/mannixp/SpectralDoubleDiffusiveConvection.git}{\texttt{github}}. 

We found spatially localised states in doubly diffusive convection in a spherical shell in two different ways. 
The first method relies on the initialization of a Newton iteration using the eigenvectors obtained in section \ref{sec:LinearProb} to compute branches emerging from the primary bifurcation. 
We anticipate that these low-amplitude solutions will develop into localised solutions or into domain-filling states from which bifurcations to convectons can be found. 
This method does not allow us to find states that are disconnected from the conduction state so we also used time-stepping as we drove the value of the thermal Rayleigh number from values near the primary Hopf bifurcation, where we can calculate time-dependent behaviour, toward larger values, where time-independent solutions emerge. 
Throughout the remainder of this paper, we will use an extension of the notation introduced by \citep{lo2010spatially} to refer to different branches of spatially localised solutions: $L_{\ell}^{X,\, \pm}$, where the subscript $\ell$ indicates the Legendre polynomial degree to which the branch is associated, $X = A$ for anticonvectons, $X = C$ for convectons and $\pm$ differentiates the branches of solutions from the same family, as described in section \ref{sec:Solutions}.

\subsection{Symmetric localised states}\label{sec:even_results}

\begin{figure}
    \centering
    \includegraphics[width=0.95\textwidth]{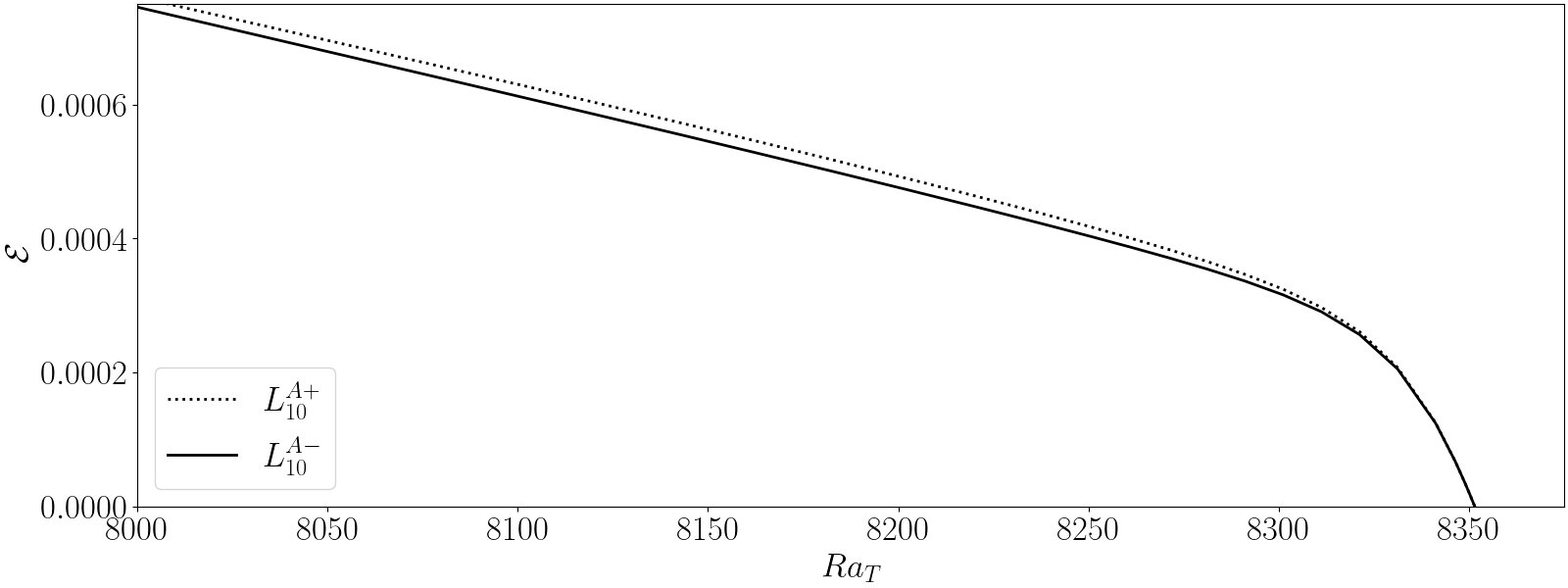}
    \caption{Bifurcation from the conduction state giving rise to the branches of anticonvectons $L_{10}^{A+}$ and $L_{10}^{A-}$. The bifurcation diagram shows the volume averaged kinetic energy $\mathcal{E}=  \frac{1}{2V} \int_{-\pi}^{\pi} \int_{r_1}^{r_2} |\boldsymbol{u}|^2 \, r^2 \sin \theta \, dr \, d \theta$ as a function of the thermal Rayleigh number $Ra_T$ for $Ra_S = 400$ and $\Gamma = 8.9224$. As explained in the text, the presence of a hidden symmetry implies that this bifurcation is a codimension-2 bifurcation between a transcritical and a saddle-node bifurcation.} 
    \label{fig:Transcritical}
\end{figure}

\begin{figure}
    \centering
    \includegraphics[width=\textwidth]{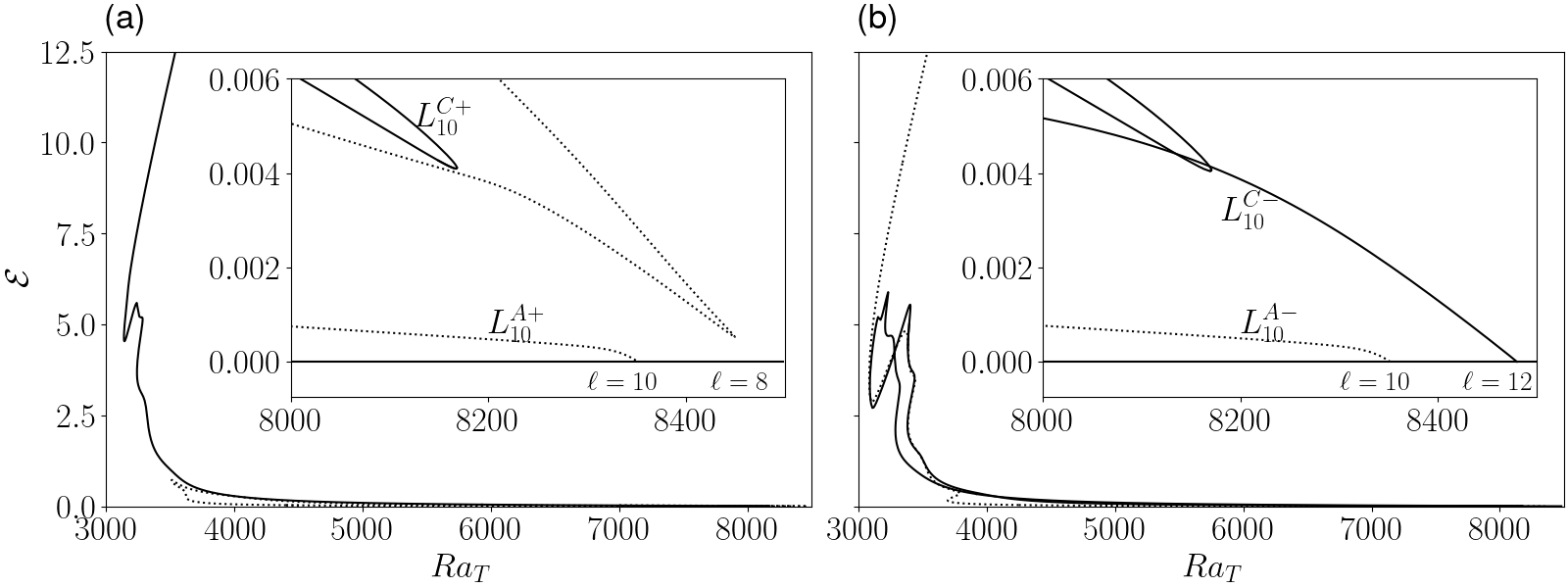}
    \caption{Bifurcation diagram of the $L_{10}^{A+}$ anticonvecton and $L_{10}^{C+}$ equatorial-convecton branches (a) and of the $L_{10}^{A-}$ anticonvecton and $L_{10}^{C-}$ equatorial-convecton branches (b). The equatorial-convectons (resp. anticonvectons) are indicated by a solid (resp. dotted) line. Parameters are $Ra_S = 400$ and $\Gamma = 8.9224$.}
    \label{fig:full_bif_even_parity}
\end{figure}

Figure \ref{fig:Transcritical} shows the different branches of solutions which emerge at the steady state bifurcation generated by the $\ell=10$ mode at $Ra_T = 8351.53$ for $Ra_S = 400$ and $\Gamma = 8.9224$.
As this bifurcation does not break the $R_{AP}$ symmetry, one would expect it to produce two branches of solutions, one supercritical and one subcritical. 
As explained in section \ref{sec:LinearProb}, this bifurcation is, in fact, a TSN bifurcation where the saddle-node of the supercritically emerging branch occurs at the same location as the transcritical bifurcation from the conduction state \citep{beltrame2015onset}. 
The resulting TSN bifurcation produces two subcritical branches bearing symmetric solutions: $L_{10}^{A+}$ anticonvectons and $L_{10}^{A-}$ anticonvectons. 
Figure \ref{fig:full_bif_even_parity} shows the path of these branches as they are followed either until their termination point or until they reach large amplitude states.
Starting with $L_{10}^{A+}$, as the branch is continued to lower $Ra_T$, the solution becomes spatially localised (see figure \ref{fig:Intro_Even_Parity}(c)). 
However, it does not produce any snaking like that of figure \ref{fig:planar}: the branch quickly turns around to larger $Ra_T$ and smaller energy before turning around near the $\ell = 8$ bifurcation point. 
The initial stages of the bifurcation scenario of the $L_{10}^{A-}$ branch are similar and it produces anticonvectons with an upflow at the poles (see figure \ref{fig:Intro_Even_Parity}(d)). 
Unlike the $L_{10}^{A+}$ branch, however, the $L_{10}^{A-}$ branch goes on to larger energies while producing a behaviour akin to snaking. 
For the sake of simplicity, we will hereafter use the word snaking loosely to refer to the behaviour of solution branches that undergo series of saddle-node bifurcations or oscillations in parameter space. 
By using time-stepping at $Ra_T = 4000$, we were able to compute a stable large amplitude equatorial-convecton in its domain-filling form. 
We numerically continued it to trace the $L_{10}^{C+}$ branch in figure \ref{fig:full_bif_even_parity}(a). 
This branch produces snaking but does not connect the conduction state. 
Instead, it turns around a saddle-node at low amplitude before undergoing complex behaviour (not shown). 
We were also able to compute another type of equatorial-convectons: $L_{10}^{C-}$ (see figure \ref{fig:full_bif_even_parity}(b)). 
These convectons produce snaking branches but, instead of extending to large amplitude like $L_{10}^{C+}$, they turn back to low amplitude where they connect with the conduction state at the $\ell = 12$ bifurcation. 
The behaviour of these localised state branches is not trivial and hints at the presence of imperfect bifurcations, especially in the case of $L_{10}^{A+}$ and $L_{10}^{C-}$. 

The bifurcation scenario presented here differs from the one observed in a planar geometry with periodic boundary conditions, where the primary bifurcation is triggered by eigenmodes with discrete translational invariance and where localised states only emerge from a secondary bifurcation that modulates the amplitude of the solution \citep{Beaume2011}.
Owing to the spherical geometry, the eigenfunctions are spatially modulated and favour the formation of convection rolls near the poles. 
Due to this spatial modulation, the aforementioned secondary bifurcation cannot be achieved and the branch of spatially localised states with convection rolls near the poles, $L_{10}^{A \pm}$, consequently bifurcates directly from the conduction state (as was observed for the Swift--Hohenberg equation on the axisymmetric disc by \cite{Lloyd09, verschueren_localized_2021}). 
The branches of spatially localised states with convection rolls near the equator, $L_{10}^{C \pm}$, on the other hand, are disconnected from the branch bifurcating from the $\ell=10$ mode. 
We attribute this disconnection to the fact that the spatial modulation of equatorial convectons is opposite to that of the eigenfunctions (i.e., the Legendre polynomials shown in section \ref{sec:LinearProb}).

As the spherical problem yields four distinct branches of symmetric localised states which would be dynamically equivalent if the system were translational invariant in $\theta$, strong similarities are observed with the bifurcation structure of a planar system in the absence of this symmetry \citep{Houghton2009, mercader2009localized}.
To shed further light on the bifurcation structure, we treat the solutal Rayleigh number, $Ra_S$, as a homotopy parameter, i.e., we track how the bifurcation diagram changes with $Ra_S$. 
We describe our findings below, together with a description of the solutions belonging to the various localised state branches computed.

\subsubsection{Minus branch}

\begin{figure}
    \centering
    \includegraphics[width=\textwidth]{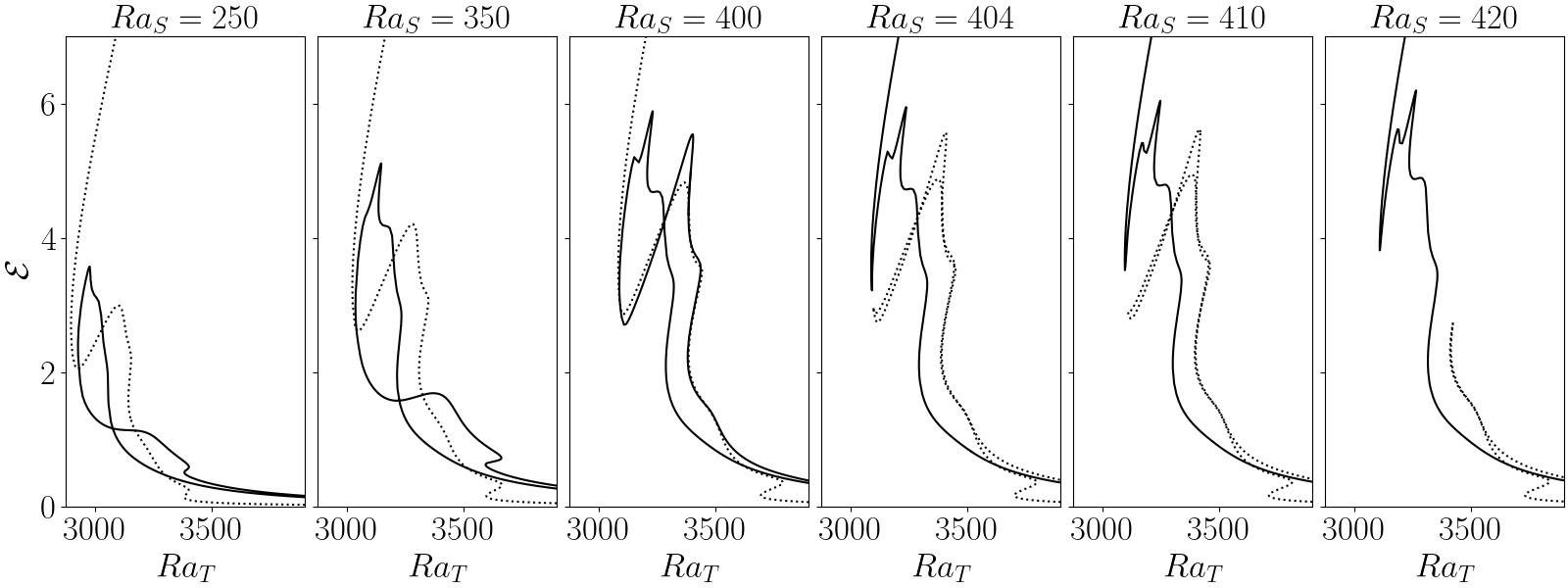}
    \caption{Bifurcation diagrams of the $L_{10}^{A-}$ anticonvecton (dotted line) and the $L_{10}^{C-}$ equatorial-convecton (solid line) branches shown by the kinetic energy $\mathcal{E}$ as a function of the thermal Rayleigh number $Ra_T$ for different values of $Ra_S$.}
    \label{fig:bif_l10m_compare}
\end{figure}

Figure \ref{fig:bif_l10m_compare} shows how the bifurcation diagram of the $L_{10}^{A-}$ and $L_{10}^{C-}$ branches changes with the value of the solutal Rayleigh number.
An imperfect bifurcation is present at $Ra_S \approx 404$ that splits the equatorial-convectons from the anticonvectons at their leftmost saddle-node, the point where these branches would otherwise connect spatially periodic states in suitable domains \citep{Houghton2009}. 
For larger $Ra_S$, the $L_{10}^{C-}$ equatorial-convectons are favoured over the $L_{10}^{A-}$ anticonvectons whose extent decreases rapidly while the equatorial-convecton branch remains the only one present at large energies. 
When $Ra_S$ is decreased below the imperfect bifurcation, the opposite behaviour is observed, although the process is much slower: the equatorial-convectons at $Ra_S = 250$ still reach larger energies than the anticonvectons at $Ra_S = 420$. 
Changing the value of the solutal Rayleigh number also affects the minimum value of the thermal Rayleigh number at which these localised states can be observed: the leftmost saddle-node of the dominant structure is found at $Ra_T \approx 2896.72$ for $Ra_S = 250$ while it is located at $Ra_T \approx 3110$ for $Ra_S = 420$.

\begin{figure}
    \centering
    \includegraphics[width=0.95\textwidth]{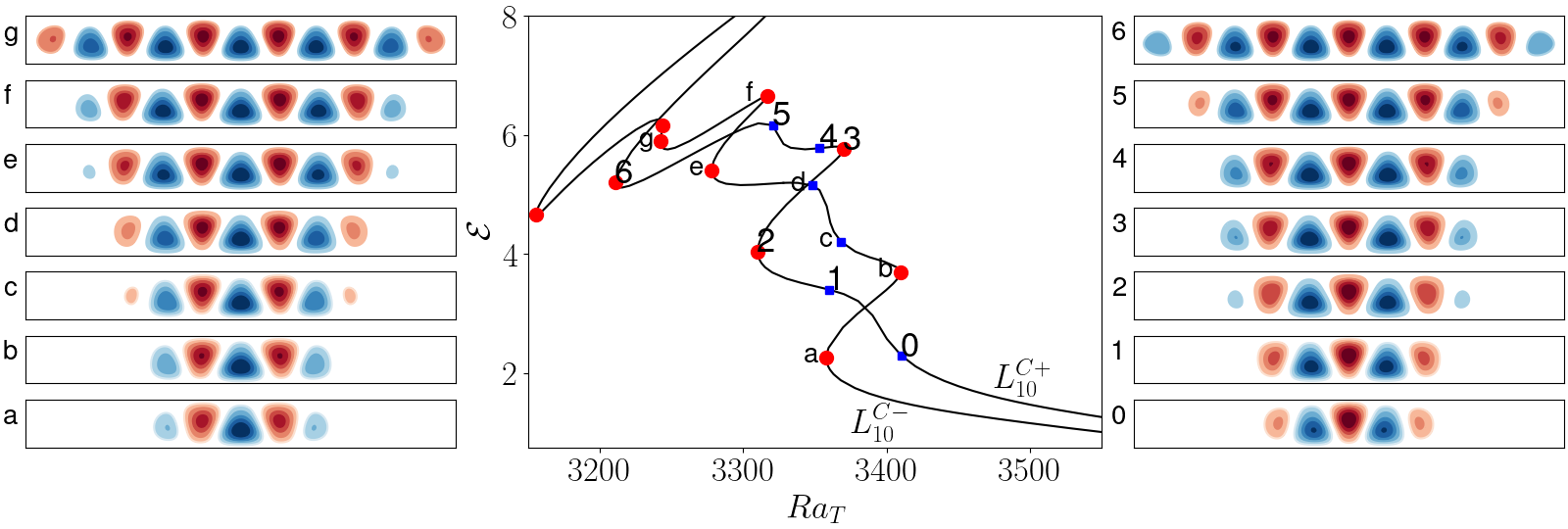}
    \caption{(Middle) Bifurcation diagram showing the $L_{10}^{C-}$ and the $L_{10}^{C+}$ equatorial-convecton branches via the kinetic energy $\mathcal{E}$ as a function of the thermal Rayleigh number $Ra_T$. Solutions at the saddle-nodes (red dots) and at additional points (filled blue squares) along $L_{10}^{C-}$ are represented on the left panel and along $L_{10}^{C+}$ on the right panel. These representations show the temperature departure from the conduction state profile $\Theta$. The solutal Rayleigh number value is $Ra_S=450$.}
    \label{fig:l10m_convectons}
\end{figure}

By looking at both sides of the imperfect bifurcation identified in figure \ref{fig:bif_l10m_compare}, we were also able to compute the full extent of the $L_{10}^{A-}$ anticonvectons and of the $L_{10}^{C-}$ equatorial-convecton spatial development.
The $L_{10}^{C-}$ equatorial-convectons are represented together with their bifurcation diagram at $Ra_S = 450$ in figure \ref{fig:l10m_convectons}.
Starting from low energy, the branch first extends to lower thermal Rayleigh numbers where the solution takes the form of an equatorial-convecton displaying downflow at the equator and constituted of $4$ convection rolls, leading to downward flow at both ends of the localised structure. 
This structure is shown in solution \textbf{a} in figure \ref{fig:l10m_convectons}, where $5$ adjacent regions are visible where the temperature departure from the conduction state $\Theta$ alternates sign.
Following this branch to larger $\mathcal{E}$, we observe that the $L_{10}^{C-}$ equatorial-convectons grow in a similar way to snaking localised states with two important exceptions: (i) the branch oscillation in parameter space is of smaller amplitude when the outermost rolls display upflow on the outside of the convective structure and (ii) the snaking branch is slanted to the left and the saddle-nodes are, therefore, not aligned.
Nonetheless, the solution grows in space by successive nucleation of convection rolls after every oscillation of the branch.
In the case of $L_{10}^{C-}$ at $Ra_S = 450$, the slant is such that some saddle-nodes are missing thus preventing a one-to-one comparison with standard snaking (see figure \ref{fig:planar}). 
To complete the picture of the nucleation process, we manually added points \textbf{c} and \textbf{d} on the $L_{10}^{C-}$ branch of figure \ref{fig:l10m_convectons} which would roughly correspond to the location of saddle-nodes if the branch were not slanted.
The equatorial-convecton becomes domain-filling at point \textbf{g} before adjusting its wavelength to the size of the domain and leaving the slanted snaking region to go to large amplitude and large thermal Rayleigh number.

\begin{figure}
    \centering
    \includegraphics[width=0.63\textwidth]{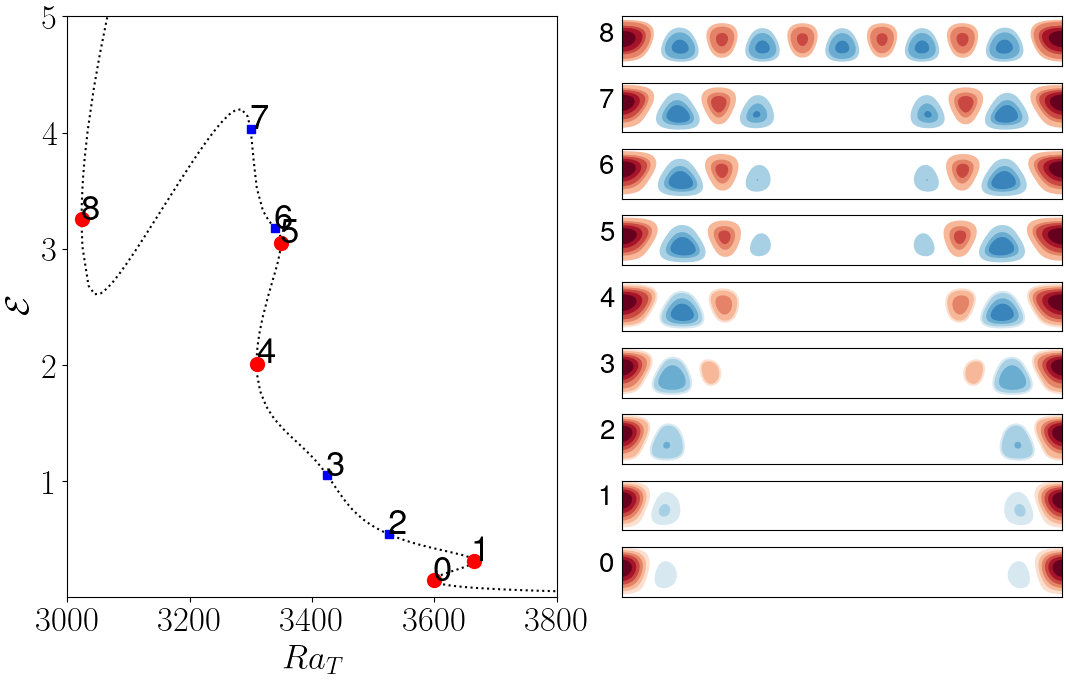}
    \caption{(Left) Bifurcation diagram showing the $L_{10}^{A-}$ anticonvecton branch via the kinetic energy $\mathcal{E}$ as a function of the thermal Rayleigh number $Ra_T$. (Right) Solutions at the saddle-nodes (red dots on the left panel) and at additional points (blue dots on the left panel) along the branch represented by the temperature departure from the conduction state profile $\Theta$. The solutal Rayleigh number value is $Ra_S=350$.}
    \label{fig:l10m_anti_convectons}
\end{figure}

The $L_{10}^{A-}$ anticonvecton branch is represented at $Ra_S = 350$ in figure \ref{fig:l10m_anti_convectons}.
At its lowest-energy saddle-node, the $L_{10}^{A-}$ anticonvectons take the form of identical localised convection states located at both poles comprised of one roll driving an upward flow at the pole. 
The branch proceeds in a similar fashion to $L_{10}^{C-}$ at $Ra_S = 450$, describing a similar type of slanted imperfect snaking. 
The $L_{10}^{A-}$ anticonvectons also grow similarly: one convection roll is added around each polar structure per branch oscillation until a domain-filling, $10$-roll state is reached by saddle-node \textbf{8}. 
The branch then turns around to larger thermal Rayleigh numbers while convection strengthens. 
This spatially developed convection state connects with the one produced by the $L_{10}^{C-}$ branch at the imperfect bifurcation at $Ra_S \approx 404$. 

\subsubsection{Plus branch}

\begin{figure}
    \centering
    \includegraphics[width=0.99\textwidth]{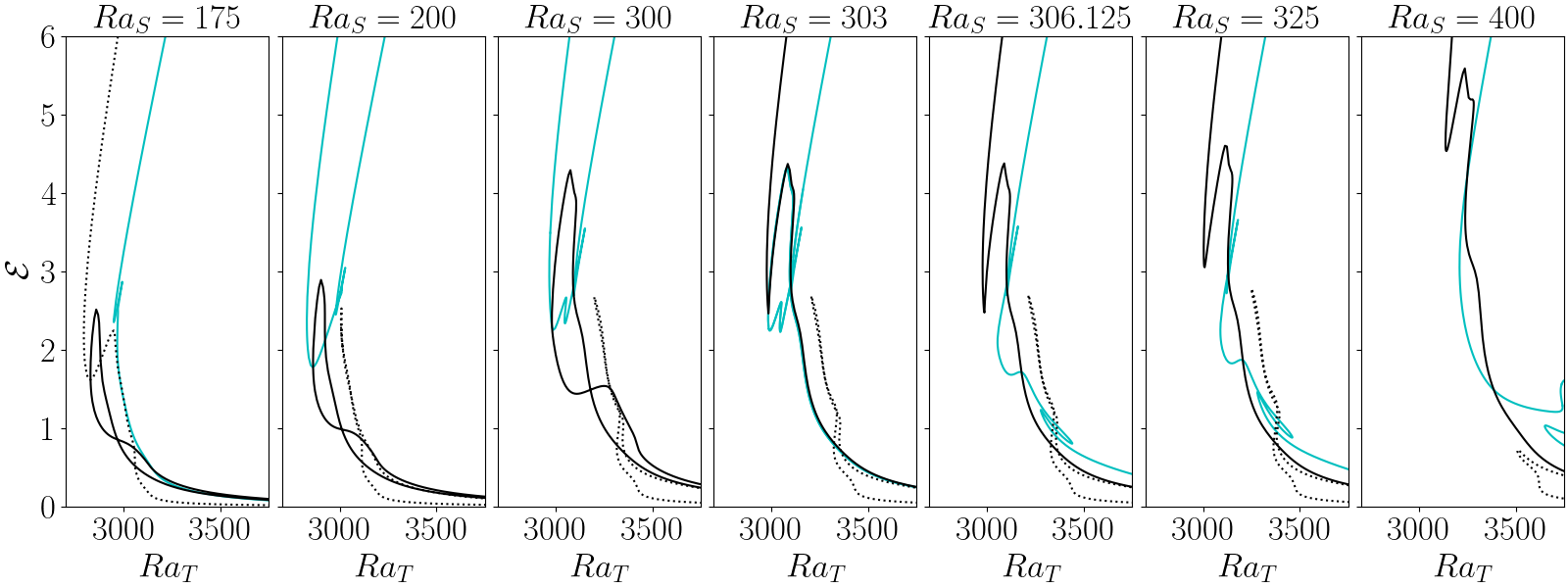}
    \caption{Bifurcation diagram of the $L_{10}^{A+}$ anticonvecton (dotted line) and the $L_{10}^{C+}$ equatorial-convecton (solid line) branches shown by the kinetic energy $\mathcal{E}$ as a function of the thermal Rayleigh number $Ra_T$ for different values of $Ra_S$. The branch shown in blue is shown for completeness as it is involved in an imperfect bifurcation with $L_{10}^{A+}$ for $Ra_S \approxeq 198$.}
    \label{fig:bif_l10p_compare}
\end{figure}

\begin{figure}
    \centering
    \includegraphics[width=0.63\textwidth]{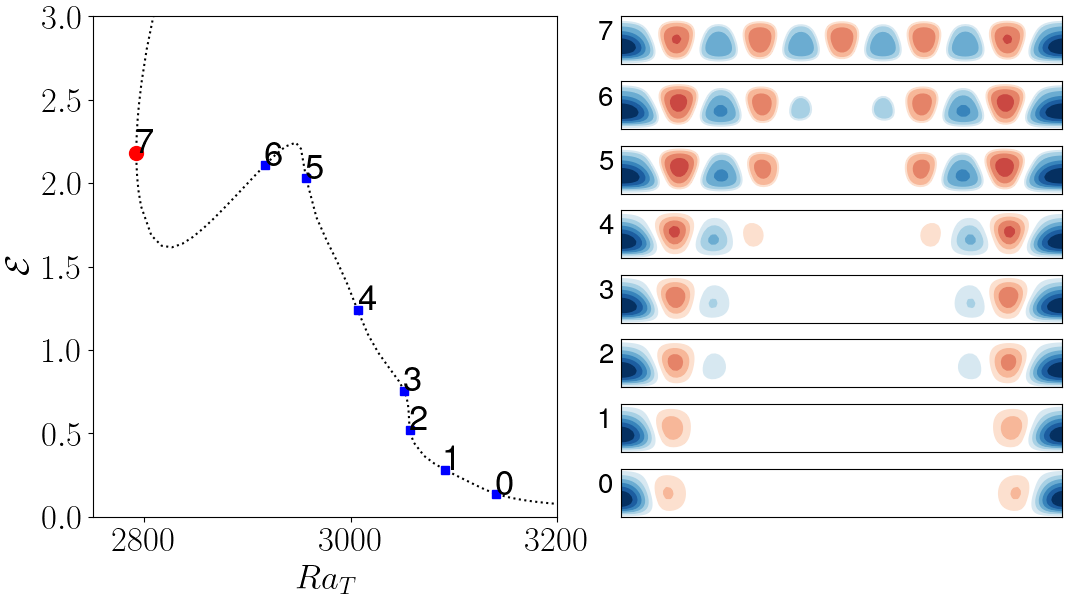}
    \caption{(Left) Bifurcation diagram showing the $L_{10}^{A+}$ anticonvecton branch via the kinetic energy $\mathcal{E}$ as a function of the thermal Rayleigh number $Ra_T$. (Right) Solutions at the saddle-nodes (red dots on the left panel) and at additional points (blue dots on the left panel) along the branch represented by the temperature departure from the conduction profile $\Theta$. The solutal Rayleigh number value is $Ra_S = 175$.}
    \label{fig:l10p_anti_convectons}
\end{figure}

The scenario associated with the $L_{10}^{A+}$ and $L_{10}^{C+}$ branches can also be elucidated by using the solutal Rayleigh number as a homotopy parameter.
Figure \ref{fig:bif_l10p_compare} shows how these branches behave for several values of $Ra_S$.
The connection scenario between $L_{10}^{A+}$ and $L_{10}^{C+}$ is not as direct as that between $L_{10}^{A-}$ and $L_{10}^{C-}$. 
It involves two imperfect bifurcations and a third branch of solutions. 
At $Ra_S = 400$, the $L_{10}^{C+}$ equatorial-convectons extend to large energy while the $L_{10}^{A+}$ anticonvectons turn around at a saddle-node and remain low energy states. 
A first imperfect bifurcation takes place at $Ra_T \approx 303$ where the upper part of the $L_{10}^{C+}$ branch disconnects into a branch of domain-filling states, leaving the spatially localised states on a branch that returns back to low energy.
As $Ra_S$ is reduced, the branch of equatorial-convectons recedes while the opposite happens to the branch of anticonvectons. 
A second imperfect bifurcation takes place at $Ra_S \approx 198$, where the anticonvecton branch connects with a branch of domain-filling states with $12$ rolls and allows the $L_{10}^{A+}$ anticonvectons to extend to high energy at lower values of the solutal Rayleigh number. 
The $L_{10}^{C+}$ equatorial-convectons are best described at larger values of the solutal Rayleigh number. 
They are represented at $Ra_S = 450$ in figure \ref{fig:l10m_convectons} and behave in a similar way to the $L_{10}^{C-}$ equatorial-convectons with the exception that they drive an upward rather than a downward flow at the equator. 
Conversely, the $L_{10}^{A+}$ anticonvectons are best described at lower values of the solutal Rayleigh number and are shown in figure \ref{fig:l10p_anti_convectons} for $Ra_S = 175$.
Although the $L_{10}^{A+}$ anticonvectons grow in a similar fashion to their $L_{10}^{A-}$ counterparts, the oscillations of their branch as it is followed toward larger energy is not as pronounced as those of $L_{10}^{A-}$. 
Additionally, we find that the convective regions are not as strongly localised, with small oscillations of the temperature field remaining in the quiescent regions of the localised states of figure \ref{fig:l10p_anti_convectons}.
We attribute both of these differences to the lower value of the solutal Rayleigh number, which implies a larger disparity between the solutal and thermal buoyancy effects and, therefore, a weaker subcriticality.

An important feature of the bifurcation scenario presented here and which is different in planar geometries, is the fact that full branches of equatorial-convectons are found at larger values of the solutal Rayleigh number while full branches of anticonvectons are found at lower values of the same parameter.
This selection mechanism may be related to the balance of contributions to the buoyancy force, usually characterised by the magnitude of the buoyancy ratio: $|N| = Ra_S/Ra_T$ (for $Ra_S \leq Ra_T$). 
The further away from $1$ the value of this number, the less competition there is between thermal and solutal effects, and the weaker the subcriticality is expected to be.
Anticonvectons are found for relatively low buoyancy ratio magnitudes: for example the full $L_{10}^{A-}$ branch was computed for $|N| \approx 0.10$ in figure \ref{fig:l10m_anti_convectons} while the full $L_{10}^{A+}$ branch was computed for $|N| \approx 0.06$ in figure \ref{fig:l10p_anti_convectons}. 
These solutions are characterised by localised convection near the poles. 
This is where the geometry favours the emergence of convection, as shown by the eigenmodes of the linearised problem (see figure \ref{fig:EigenVectors}).
Conversely, full branches of equatorial-convectons of figure \ref{fig:l10m_convectons} are computed for larger buoyancy ratio magnitudes, $|N| \approx 0.14$ and are not found for $|N| < 0.10$. 
We attribute this to the fact that, unlike for the anticonvectons, their existence is not facilitated by geometrical properties and, thus, relies critically on the balance between thermal and solutal effects to sustain convection. 

The snaking observed for both equatorial-convectons and anticonvectons is also characterised by the fact that, when the outermost rolls display downward flow (i.e., a negative temperature perturbation) on the outside of the convective structure, the branch covers a wider range of thermal Rayleigh number values than when these rolls display upward flow (i.e., a positive temperature perturbation).
We attribute this behaviour to the fact that the spherical shell system, unlike the planar one, does not possess any top-down symmetry. 
Because of this, the nucleation of new rolls is a different process depending on its sense of rotation. 
A roll yielding downward flow at the limit of the convective structure is associated with the growth of a negative temperature perturbation. 
In our figures, these perturbations take the form of blue trapezia with a smaller upper side. 
Conversely, nucleation of the opposite rolls creates a positive temperature perturbation which looks like red trapezia with a longer upper side.

\subsection{Symmetry-breaking localised states}\label{sec:odd_results}

In addition to the symmetric equatorial-convectons and anticonvectons, system \eqref{eq:3D_EQNS} admits spatially localised solutions that break the equatorial symmetry $R_{AP}$. 
These states may originate from pitchfork bifurcations from the conduction state which are generated by marginal eigenmodes of the linear stability analysis with odd parity. 
Unlike $R_{AP}$-symmetric solutions, which we sought on the symmetric (i.e., reduced), subspace, symmetry-breaking localised states necessitate the whole search space to be computed.

\begin{figure}
    \centering
    \includegraphics[width=.95\textwidth]{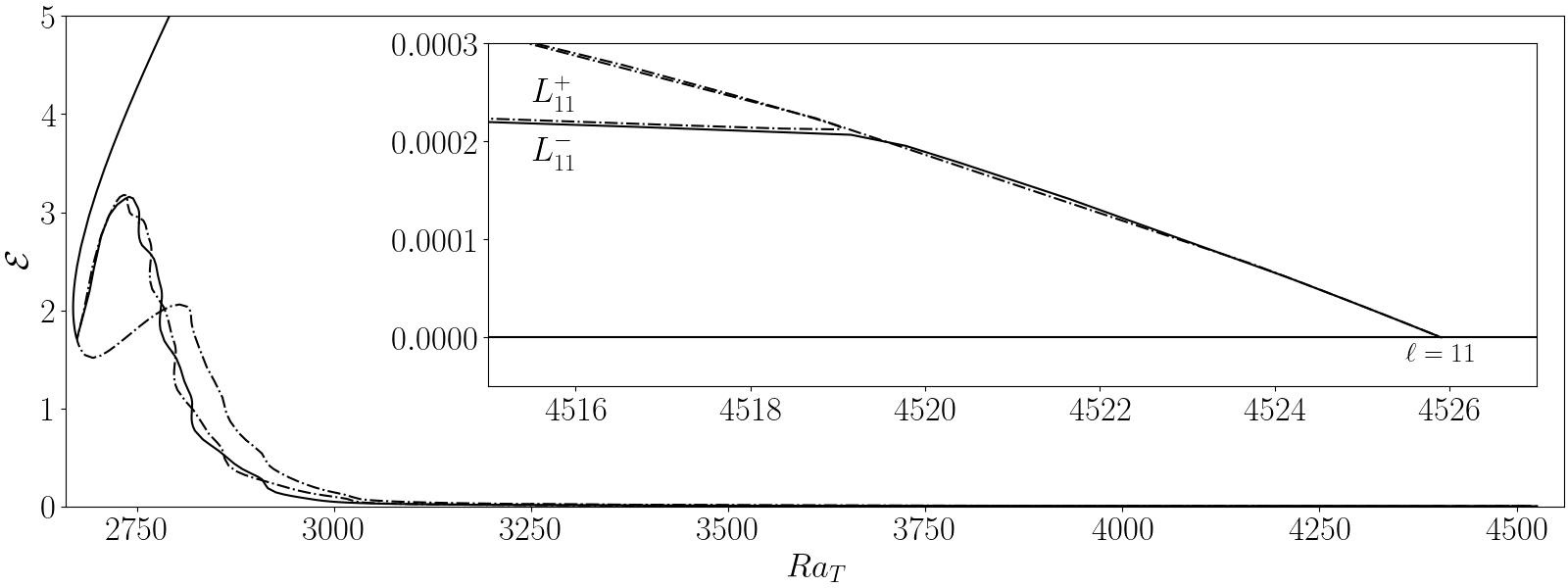}
     \caption{Bifurcation diagram showing the kinetic energy $\mathcal{E}$ as a function of the thermal Rayleigh number $Ra_T$ for the $L^+_{11}$ (dashed) and $L^-_{11}$ (solid) pole-convecton branches. Parameters are $Ra_S=150$ and $\Gamma = 10.029$.}
    \label{fig:full_bif_odd_parity}
\end{figure}

\begin{figure}
    \centering
    \includegraphics[width=0.995\textwidth]{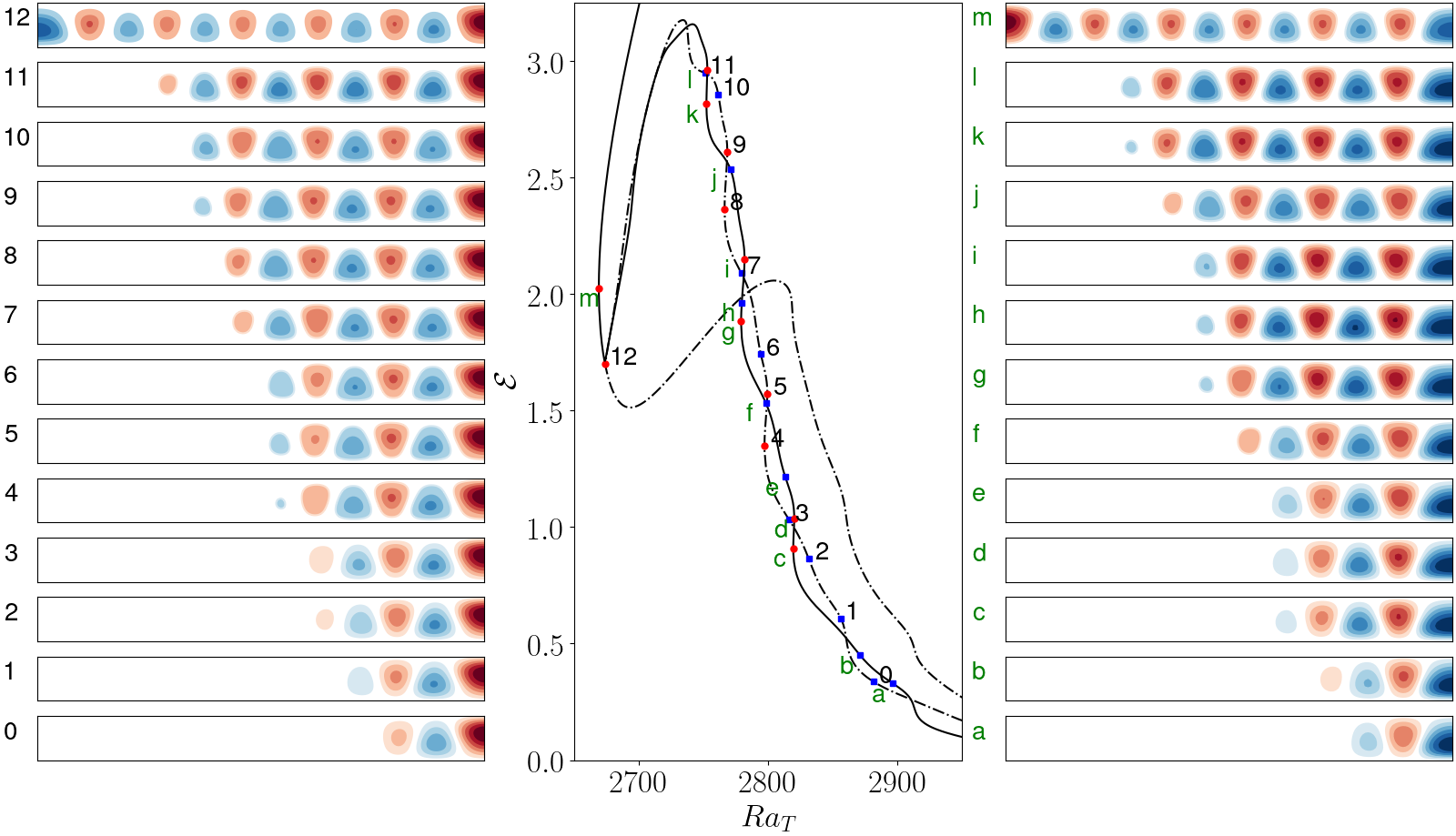}
    \caption{Bifurcation diagram of the snaking region (middle) of the  $L^-_{11}$ pole-convectons (solid line, right) and of the $L^+_{11}$ pole-convectons (dashed line, left) showing the solution kinetic energy $\mathcal{E}$ as a function of the thermal Rayleigh number $Ra_T$. Solutions are taken at successive saddle nodes or roll nucleation points and represented in the left and right panels via the temperature departure from conduction $\Theta$. Parameter values are $Ra_S = 150$ and $\Gamma = 10.029$. }
    \label{fig:l11_snaking}
\end{figure}

\begin{figure}
    \centering
    \includegraphics[width=0.66\textwidth]{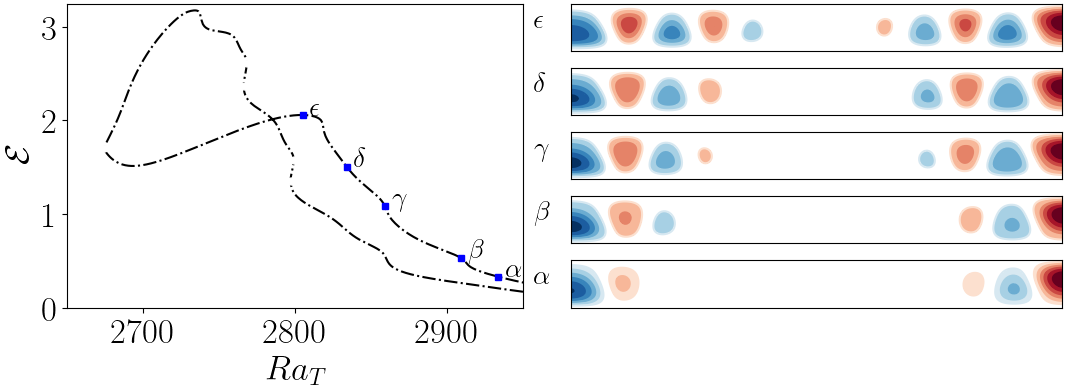}
    \caption{(Left) Bifurcation diagram showing the $L^+_{11}$ branch of figure \ref{fig:l11_snaking} (dashed line) using the kinetic energy $\mathcal{E}$ as a function of the thermal Rayleigh number $Ra_T$. (Right) Solutions are taken at the points labelled on the left panel and represented by the temperature departure from conduction $\Theta$. Parameter values are $Ra_S = 150$ and $\Gamma = 10.029$. Solutions on the other part of the branch are shown in figure \ref{fig:l11_snaking}.}
    \label{fig:l11_snaking_extra_detail}
\end{figure}

The complete bifurcation diagram of the $R_{AP}$-breaking localised states produced by the pitchfork bifurcation at $\ell = 11$ is shown in figure \ref{fig:full_bif_odd_parity} for $Ra_S = 150$.
The $\ell = 11$ pitchfork at $Ra_T \approx 4526$ produces subcritical branches of domain-filling counterrotating rolls of convection that reach an imperfect bifurcation point at $Ra_T \approx 4519$. 
Following the branch whose energy grows the slowest past the imperfect bifurcation, spatial modulation develops to favour the pole where the structure displays downflow. 
The resulting solutions are referred to as $L_{11}^-$ pole-convectons and the corresponding branch extends subcritically to produce slanted snaking between $Ra_T = 2750$ and $Ra_T = 3000$, described further in figure \ref{fig:l11_snaking}.
This time, the snaking produced by the $L_{11}^-$ branch appears more regular.
With the exception of the bottom saddle-nodes, which may not always align with the others \citep{chapman09}, the saddle-nodes appear to fall within a snaking region that is slightly slanted to the left. 
As the branch is followed upward along the snaking, typical spatial growth is observed: a counterrotating roll of convection is nucleated on the outside of the convective structure during each snaking period until the domain is full.
Just like for the $R_{AP}$-symmetric equatorial-convectons and anticonvectons, we observe that the nucleation of a roll driving a downward flow at the edge of the convective structure leads to a wider amplitude oscillation of the branch in parameter space than the nucleation of a roll associated with an upward flow at the edge of the convective structure. 
At what would be its termination point, where the pattern fills the domain, the branch undergoes another imperfect bifurcation induced by the domain geometry at $Ra_T \approx 2674$: it turns to higher energy states, passing through saddle-node \textbf{m} and taking on the role of the domain-filling state branch.
The other side of this imperfect bifurcation contains localised states of the opposite phase, i.e., pole-convectons with an upflow at the pole, which we refer to as $L_{11}^+$.
Following $L_{11}^+$ on the side of this imperfect bifurcation where its solution energy increases, one obtains a complementary snaking to the one described by $L_{11}^-$, with a similar saddle-node alignment and spatial growth mechanism.
The branch can be followed all the way down the snaking and toward larger thermal Rayleigh numbers to reveal that it is also involved in the imperfect bifurcation at $Ra_T \approx 4519$, where it abruptly turns around toward larger amplitude states. 
The $L_{11}^+$ branch behaviour past that point is complex and beyond the scope of this paper. 
Following $L_{11}^+$ in the other direction past the imperfect bifurcation at $Ra_T \approx 2674$ reveals that the solution undergoes spatial modulation to preserve convection at both poles but with opposite flow direction.
As a result, the $L_{11}^{+}$ pole-convecton turns into  symmetry-breaking anticonvectons, as shown in figure \ref{fig:l11_snaking_extra_detail} and
the branch produces a similar bifurcation diagram to the one of the symmetric anticonvectons.
After this spatially localised interlude, the branch travels to low-amplitude states where it reconnects to the $\ell = 11$ bifurcation.

We have regretfully not been able to compute the bifurcation diagram of symmetry-breaking equatorial convectons (the symmetry-breaking analogue to the anticonvections shown in figure \ref{fig:l11_snaking_extra_detail}).
We are convinced that these states exist despite our best attempts failing to locate them.
We have tried using combinations of time-stepping, homotopical continuation and aggressive manipulations of the solution space, among other techniques.
The difficulties that prevented us from computing these solutions are: (i) their asymmetric nature, (ii) the fact that their solution branch is not connected to the conduction state, (iii) that the structure of their convection rolls is not as simple as in the planar system; and (iv) that the pole-convectons that share parity with the symmetry-breaking equatorial convectons do not possess an equatorially-centred amplitude envelope in their domain-filling form (the amplitude dip of states \textbf{12} and \textbf{m} of figure \ref{fig:l11_snaking} is located closer to the left of the domain/South pole). 
We do not, however, expect these solutions to produce substantially different behaviour from the other localised solutions presented in this paper.

\section{Discussion}\label{sec:Discussion}

Motivated by the recent progress made on spatially localised pattern formation in planar fluid systems and, in particular, in doubly diffusive convection, we focused here on an alternate geometry relevant to astrophysical applications: the spherical shell. 
In order to connect our findings in the spherical shell with established literature on spatial localisation, we chose parameter values used in studies of doubly diffusive convection in planar geometries. 
As the present work represents the first analysis of spatially localised states in a spherical shell, we restricted ourselves to axisymmetric solutions.

The most distinguishing feature of doubly diffusive convection in an axisymmetric spherical shell is the fact that convection rolls are not straight as in two-dimensional planar studies but are curved in order to wrap around the inner sphere. 
The solutions bifurcating from the conduction state are therefore different from those of the planar problem, where they consist of homogeneous arrays of identical convection rolls typically referred to as spatially periodic. 
In the spherical shell, they consist of spatially modulated solutions which become spatially localised as their solution branches are continued away from the bifurcation point.
Such solutions only arise via secondary bifurcations in unbounded planar doubly diffusive convection \citep{Beaume2011}. 
The absence of a primary bifurcation to spatially periodic states is a consequence of the fact that the axisymmetric spherical shell system is not invariant under spatial translations in the latitudinal direction. 
The fact that spatially periodic states are not created by the primary bifurcation has already been observed in systems with non-periodic boundary conditions and, therefore, in the absence of translation invariance, like in the Swift--Hohenberg equation \citep{Houghton2009}, in binary fluid convection \citep{mercader2011convectons} and in natural doubly diffusive convection \citep{beaume2013nonsnaking}. 

By performing numerical continuation, we unravelled the existence of a number of spatially localised state families, which we differentiate by their symmetry and by the location of the localised pattern, i.e., either/both pole(s) or the equator. 
In particular, we found: equatorial-convectons, where convection is centred around the equator; pole-convectons, where it is centred around one pole; and anticonvectons, where each pole is the centre of a convection zone. 
All these solutions have analogues in the planar doubly diffusive convection where they exist within similar values of the parameters \citep{mercader2011convectons,Beaume2011}.
This is not the case in the spherical shell, where spatially localised convection is preferred at the poles for relatively small values of the solutal Rayleigh number and at the equator for larger values of the same number. 
This suggests that the balance between the thermal and solutal components of the buoyancy force plays a selection role in the type of localised pattern that develops in the system.

The best theoretical framework for the analysis of doubly diffusive pole convectons is the axisymmetric Swift--Hohenberg equation (SHE) posed on a disc \citep{Lloyd09,mccalla_snaking_2010}. 
The disc on which the SHE is posed corresponds to a hemishell centred on either pole in such a way that pole-convectons (and anticonvectons) correspond to spot patterns and equatorial-convectons correspond to target patterns in the SHE.
The bifurcation diagram of axisymmetric states computed by \citep{verschueren_localized_2021} for the SHE on a disc reveals a number of similarities with our bifurcation diagrams. 
In particular, spot patterns bifurcate directly from the trivial solution while target patterns are disconnected from it.
We observed the same in doubly diffusive convection, where pole-convectons and anticonvectons are found to connect with the conduction state while equatorial-convectons are not always connected with it. 
The bifurcation diagram of the localised states in the axisymmetric SHE displays snaking that is not straight but tends to have a more pronounced leftward slant when the edge of the localised pattern approaches the centre of the domain, where the roll curvature becomes important.
While our domain size is modest by comparison with that used in standard SHE studies, these results suggests that the leftward slanted snaking that we observe is a direct consequence of the geometry.

We anticipate that the bifurcation scenario of the fully three-dimensional system will support an even larger variety of spatially localised states. 
In particular, new types of convectons may arise that break the longitudinal-invariance. 
Predictions about their formation and bifurcation diagram can be made in light of the non-axisymmetric solutions found in the Swift--Hohenberg equation by \citep{verschueren_localized_2021}. 
However, the expected complexity in the fluid system is best approached with knowledge of the axisymmetric bifurcation scenario.
This is what we have aimed to achieve here.

Besides providing here the first study of doubly diffusive convectons in a spherical shell, we believe that the scenario presented and the links made with the axisymmetric Swift--Hohenberg equation possess a wide applicability to spatially localised pattern formation in spherical geometries, in fluid dynamics and beyond.
Further work should include the computation of the symmetry-breaking equatorial-convectons that we have not been able to compute here. 
We will also endeavour to extend our findings to larger values of the domain aspect-ratio $\Gamma$, which will allow a better connection to the planar geometry literature and to Swift--Hohenberg theory.

\section*{Acknowledgements}
The authors wish to thank Jonathan Mestel for their careful reading and recommendations for improving this manuscript. The authors would also like to thank Rainer Hollerbach for their involvement in the early stages of this manuscript. P.M. Mannix was supported by the Engineering and Physical Sciences Research Council [grant number EP/V033883/1] as part of the [D$^{*}$]stratify project. 

\section*{Data access}
The numerical code, supplementary material detailing its implementation and data used to generate the figures are available on Zenodo at \url{https://doi.org/10.5281/zenodo.15462897}. 

\section*{Author ORCIDs}
Paul Mannix, https://orcid.org/0000-0003-3042-857X; \\
Cedric Beaume, https://orcid.org/0000-0003-3485-6387; 

\bibliographystyle{alpha}
\bibliography{References}

\newcommand{\etalchar}[1]{$^{#1}$}
\begin{thebibliography}{WHWM22}

\bibitem[ABM22]{alonso2022stationary}
A.~Alonso, O.~Batiste, and I.~Mercader.
\newblock Stationary localized solutions in binary convection in slightly
  inclined rectangular cells.
\newblock {\em Phys. Rev. E}, 106(5):055106, 2022.

\bibitem[BBK11]{Beaume2011}
C.~Beaume, A.~Bergeon, and E.~Knobloch.
\newblock Homoclinic snaking of localized states in doubly diffusive
  convection.
\newblock {\em Phys. Fluids}, 23, 2011.

\bibitem[BBK13]{beaume2013convectons}
C.~Beaume, A.~Bergeon, and E.~Knobloch.
\newblock Convectons and secondary snaking in three-dimensional natural doubly
  diffusive convection.
\newblock {\em Phys. Fluids}, 25(2), 2013.

\bibitem[BC15]{beltrame2015onset}
P.~Beltrame and P.~Chossat.
\newblock Onset of intermittent octahedral patterns in spherical {B}{\'e}nard
  convection.
\newblock {\em European J. Mech. B-Fluids}, 50:156--174, 2015.

\bibitem[Bea17]{beaume2017adaptive}
C.~Beaume.
\newblock Adaptive {S}tokes preconditioning for steady incompressible flows.
\newblock {\em Commun. Comput. Phys.}, 22(2):494--516, 2017.

\bibitem[BHK09]{Burke2009}
J.~Burke, S.~M. Houghton, and E.~Knobloch.
\newblock Swift--{H}ohenberg equation with broken reflection symmetry.
\newblock {\em Phys. Rev. E}, 2009.

\bibitem[BHL25]{bramburger25arxiv}
J.~J. Bramburger, D.~J. Hill, and D.~J.~B. Lloyd.
\newblock Localized patterns.
\newblock 2025.

\bibitem[BK06]{burke06}
J.~Burke and E.~Knobloch.
\newblock Localized states in the generalized {S}wift--{H}ohenberg equation.
\newblock {\em Phys. Rev. E}, 73:056211, 2006.

\bibitem[BK08]{bergeon2008periodic}
A.~Bergeon and E.~Knobloch.
\newblock Periodic and localized states in natural doubly diffusive convection.
\newblock {\em Physica D}, 237(8):1139--1150, 2008.

\bibitem[BKB13]{beaume2013nonsnaking}
C.~Beaume, E.~Knobloch, and A.~Bergeon.
\newblock Nonsnaking doubly diffusive convectons and the twist instability.
\newblock {\em Phys. Fluids}, 25(114102), 2013.

\bibitem[Bla99]{blanchflower99}
S.~Blanchflower.
\newblock Magnetohydrodynamic convectons.
\newblock {\em Phys. Lett. A}, 261:74--81, 1999.

\bibitem[BVO{\etalchar{+}}20]{burns2020dedalus}
K.~J. Burns, G.~M. Vasil, J.~S. Oishi, D.~Lecoanet, and B.~P. Brown.
\newblock Dedalus: A flexible framework for numerical simulations with spectral
  methods.
\newblock {\em Phys. Rev. Res.}, 2(2):023068, 2020.

\bibitem[CK91]{Crawford91}
J.~D. Crawford and E.~Knobloch.
\newblock Symmetry and symmetry-breaking bifurcations in fluid dynamics.
\newblock {\em Annu. Rev. Fluid Mech.}, 23(Volume 23, 1991):341--387, 1991.

\bibitem[CK09]{chapman09}
S.~J. Chapman and G.~Kozyreff.
\newblock Exponential asymptotics of localised patterns and snaking bifurcation
  diagrams.
\newblock {\em Physica D}, 238(319--354), 2009.

\bibitem[DSH09]{duguet09}
Y.~Duguet, P.~Schlatter, and D.~S. Henningson.
\newblock Localized edge states in plane {C}ouette flow.
\newblock {\em Phys. Fluids}, 21:111701, 2009.

\bibitem[ETAS94]{Edwards1994}
W.~S. Edwards, L.~S. Tuckerman, Friesner~R. A., and D.~C. Sorensen.
\newblock Krylov methods for the incompressible {N}avier--{S}tokes equations.
\newblock {\em J. Comput. Phys.}, 110:82--102, 1994.

\bibitem[FK23]{foster23}
B.~Foster and Knobloch.
\newblock Elastic fingering in a rotating hele-shaw cell.
\newblock {\em Phys. Rev. E}, 107:065104, 2023.

\bibitem[FVKG22]{foster22}
B.~Foster, N.~Verschueren, E.~Knobloch, and L.~Gordillo.
\newblock Universal wrinkling of supported elastic rings.
\newblock {\em Phys. Rev. Lett.}, 129:164301, 2022.

\bibitem[Gar18]{garaud2018double}
P.~Garaud.
\newblock Double-diffusive convection at low prandtl number.
\newblock {\em Annu. Rev. Fluid Mech.}, 50(1):275--298, 2018.

\bibitem[GM97]{ghorayeb97}
K.~Ghorayeb and A.~Mojtabi.
\newblock Double diffusive convection in a vertical rectangular cavity.
\newblock {\em Phys. Fluids}, 9:2339--2348, 1997.

\bibitem[GS16]{Gibson2016}
J.~F. Gibson and T.~M. Schneider.
\newblock Homoclinic snaking in plane {C}ouette flow: Bending, skewing and
  finite-size effects.
\newblock {\em J. Fluid Mech.}, 794:530--551, 5 2016.

\bibitem[HK09]{Houghton2009}
S.~M. Houghton and E.~Knobloch.
\newblock Homoclinic snaking in bounded domains.
\newblock {\em Phys. Rev. E}, 80, 8 2009.

\bibitem[Kno15]{knobloch2015spatial}
E.~Knobloch.
\newblock Spatial localization in dissipative systems.
\newblock {\em Annu. Rev. Condens. Mat. Phys.}, 6:325--359, 3 2015.

\bibitem[LBK10]{lo2010spatially}
D.~{Lo Jacono}, A.~Bergeon, and E.~Knobloch.
\newblock Spatially localized binary fluid convection in a porous medium.
\newblock {\em Physics of Fluids}, 22(7), 2010.

\bibitem[LBK17]{lojacono17}
D.~{Lo Jacono}, A.~Bergeon, and E.~Knobloch.
\newblock Localized traveling pulses in natural doubly diffusive convection.
\newblock {\em Phys. Rev. Fluids}, 2:093501, 2017.

\bibitem[LGRR15]{Lloyd15}
D.~J.~B. Lloyd, C.~Gollwitzer, I.~Rehberg, and R.~Richter.
\newblock Homoclinic snaking near the surface instability of a polarisable
  fluid.
\newblock {\em J. Fluid Mech.}, 783:283--305, 2015.

\bibitem[LS09]{Lloyd09}
D.~J.~B. Lloyd and B.~Sandstede.
\newblock Localized radial solutions of the {S}wift-–{H}ohenberge quation.
\newblock {\em Nonlinearity}, 22(485--524), 2009.

\bibitem[MBAK09]{mercader2009localized}
I.~Mercader, O.~Batiste, A.~Alonso, and E.~Knobloch.
\newblock Localized pinning states in closed containers: homoclinic snaking
  without bistability.
\newblock {\em Phys. Rev. E}, 80(2):025201, 2009.

\bibitem[MBAK11]{mercader2011convectons}
I.~Mercader, O.~Batiste, A.~Alonso, and E.~Knobloch.
\newblock Convectons, anticonvectons and multiconvectons in binary fluid
  convection.
\newblock {\em J. Fluid Mech.}, 667:586--606, 2011.

\bibitem[MBAK19]{mercader19}
I.~Mercader, O.~Batiste, A.~Alonso, and E.~Knobloch.
\newblock Effect of small inclination on binary convection in elongated
  rectangular cells.
\newblock {\em Phys. Rev. E}, 99:023113, 2019.

\bibitem[MS10]{mccalla_snaking_2010}
S.~McCalla and B.~Sandstede.
\newblock Snaking of radial solutions of the multi-dimensional
  {Swift}--{Hohenberg} equation: {A} numerical study.
\newblock {\em Physica D}, 239(16):1581--1592, August 2010.
\newblock Publisher: Elsevier.

\bibitem[MS13]{mccalla_spots_2013}
S.~G. McCalla and B.~Sandstede.
\newblock Spots in the {s}wift--{h}ohenberg equation.
\newblock {\em SIAM J. Appl. Dyn. Sys.}, 12(2):831--877, 2013.

\bibitem[MT87]{Marcus1987}
P.~S. Marcus and L.~S. Tuckerman.
\newblock Simulation of flow between concentric rotating spheres. {P}art 1.
  {S}teady states.
\newblock {\em J. Fluid Mech.}, 185:1--30, 1987.

\bibitem[MVCS19]{monville2019rotating}
R.~Monville, J.~Vidal, D.~C{\'e}bron, and N.~Schaeffer.
\newblock Rotating double-diffusive convection in stably stratified planetary
  cores.
\newblock {\em Geophys. J. Int.}, 219(Supplement\_1):S195--S218, 2019.

\bibitem[PBT19]{pershin19}
A.~Pershin, C.~Beaume, and S.~M. Tobias.
\newblock Dynamics of spatially localized states in transitional plane
  {C}ouette flow.
\newblock {\em J. Fluid Mech.}, 867:414--437, 2019.

\bibitem[Rad13]{radko2013double}
T.~Radko.
\newblock {\em Double-diffusive convection}.
\newblock Cambridge University Press, 2013.

\bibitem[Rin19]{rincon_dynamo_2019}
F.~Rincon.
\newblock Dynamo theories.
\newblock {\em J. Plasma Phys.}, 85(4), August 2019.
\newblock arXiv: 1903.07829 Publisher: Cambridge University Press (CUP).

\bibitem[Sch94]{schmitt94}
R.~W. Schmitt.
\newblock Double diffusion in oceanography.
\newblock {\em Annu. Rev. Fluid Mech.}, 26:255--285, 1994.

\bibitem[SGB10]{schneider2010snakes}
T.~M. Schneider, J.~F. Gibson, and J.~Burke.
\newblock Snakes and ladders: localized solutions of plane {C}ouette flow.
\newblock {\em Phys. Rev. Lett.}, 104(10):104501, 2010.

\bibitem[SLT{\etalchar{+}}15]{stoop2015curvature}
N.~Stoop, R.~Lagrange, D.~Terwagne, P.~M. Reis, and J.~Dunkel.
\newblock Curvature-induced symmetry breaking determines elastic surface
  patterns.
\newblock {\em Nature Materials}, 14(3):337--342, 2015.

\bibitem[SMT24]{skene2024nonlinear}
C.~S. Skene, F.~Marcotte, and S.~M. Tobias.
\newblock On nonlinear transitions, minimal seeds and exact solutions for the
  geodynamo.
\newblock {\em arXiv preprint arXiv:2411.05499}, 2024.

\bibitem[TBH12]{trumper2012numerical}
T.~Tr{\"u}mper, M.~Breuer, and U.~Hansen.
\newblock Numerical study on double-diffusive convection in the {E}arth’s
  core.
\newblock {\em Phys. Earth Plan. In.}, 194:55--63, 2012.

\bibitem[TBR23]{tumelty23}
J.~Tumelty, C.~Beaume, and A.~Rucklidge.
\newblock Towards convectons in the supercritical regime: Homoclinic snaking in
  natural doubly diffusive convection.
\newblock {\em SIAM J. Appl. Dyn. Sys.}, 22:1710--1742, 2023.

\bibitem[TBR25]{tumelty2025}
J.~Tumelty, C.~Beaume, and A.~M. Rucklidge.
\newblock Convectons in unbalanced natural doubly diffusive convection.
\newblock {\em Phys. Rev. Fluids}, 10:044401, 2025.

\bibitem[TLW19]{Tuckerman2019}
L.~S. Tuckerman, J.~Langham, and A.~Willis.
\newblock {\em Order-of-Magnitude Speedup for Steady States and Traveling Waves
  via {S}tokes Preconditioning in {C}hannelflow and {O}penpipeflow}, pages
  3--31.
\newblock Springer Berlin Heidelberg, 2019.

\bibitem[Tuc89]{Tuckerman1989}
L.~S. Tuckerman.
\newblock Steady-state solving via {S}tokes preconditioning; recursion
  relations for elliptic operators.
\newblock In D.~L. Dwoyer, M.~Y. Hussaini, and R.~G. Voigt, editors, {\em 11th
  International Conference on Numerical Methods in Fluid Dynamics}, pages
  573--577. Springer Berlin Heidelberg, 1989.

\bibitem[VKU21]{verschueren_localized_2021}
N.~Verschueren, E.~Knobloch, and H.~Uecker.
\newblock Localized and extended patterns in the cubic-quintic
  {Swift}--{Hohenberg} equation on a disk.
\newblock {\em Phys. Rev. E}, 104(1):014208, July 2021.
\newblock Publisher: American Physical Society.

\bibitem[WHWM22]{wong2022layering}
T.~Wong, U.~Hansen, T.~Wieseh{\"o}fer, and W.~B. McKinnon.
\newblock Layering by double-diffusive convection in the subsurface oceans of
  {E}uropa and {E}nceladus.
\newblock {\em J. Geophys. Res.: Planets}, 127(12):e2022JE007316, 2022.

\bibitem[WIN12]{watanabe2012}
T.~Watanabe, M.~Iima, and Y.~Nishiura.
\newblock Spontaneous formation of travelling localized structures and their
  asymptotic behaviour in binary fluid convection.
\newblock {\em J. Fluid Mech.}, 712:219--243, 2012.

\bibitem[WIN16]{watanabe2016}
T.~Watanabe, M.~Iima, and Y.~Nishiura.
\newblock A skeleton of collision dynamics: Hierarchical network structure
  among even-symmetric steady pulses in binary fluid convection.
\newblock {\em SIAM J. Appl. Dyn. Sys.}, 15:789--806, 2016.

\end{thebibliography}

\end{document}